\documentclass{aa}
\usepackage{graphicx}

\def\ebv{\mbox{$\rm E(B-V)$}}
\def\ms{\mbox{$\rm M_\odot$}}
\def\ds{\mbox{$\rm d_\odot$}}
\def\dgc{\mbox{$\rm d_{GC}$}}
\def\zh{\mbox{$\rm z_h$}}
\def\rd{\mbox{$\rm R_D$}}
\def\zo{\mbox{$\rm z_\odot$}}
\def\po{\mbox{$\rm\rho_\odot$}}

\begin{document}

\title{Probing disk properties with open clusters}

\author{C. Bonatto\inst{1} \and L.O. Kerber\inst{1} \and E. Bica\inst{1}
\and B.X. Santiago\inst{1}}

\offprints{C. Bonatto}

\institute{Universidade Federal do Rio Grande do Sul, Instituto de F\'\i sica, 
CP\,15051, Porto Alegre 91501-970, RS, Brazil\\
\mail{charles@if.ufrgs.br}
}

\date{Received --; accepted --}

\abstract{We use the open clusters (OCs) with known parameters available in the WEBDA 
database and in recently published papers to derive properties related to the disk
structure such as the thin-disk scale height, displacement of the Sun above the Galactic 
plane, scale length and the OC age-distribution function. The sample totals 654 OCs, 
consisting basically of Trumpler types I to III clusters whose spatial distribution 
traces out the local geometry of the Galaxy. We find that the population of OCs with ages 
younger than 200\,Myr distributes in the disk following an exponential-decay profile with 
a scale height of $\zh=48\pm3$\,pc. For the clusters with ages in the range 200\,Myr to 
1\,Gyr we derive $\zh=150\pm27$\,pc. Clusters older than 1\,Gyr distribute nearly uniformly 
in height from the plane so that no scale height can be derived from exponential fits. 
Considering clusters of all ages we obtain an average scale height of $\zh=57\pm3$\,pc. We 
confirm previous results that \zh\ increases with Galactocentric distance. The scale height 
implied by the OCs younger than 1\,Gyr outside the Solar circle is a factor $\sim1.4-2$ larger 
than \zh\ of those interior to the Solar circle. We derive the displacement of the Sun above 
the Galactic plane as $\rm\zo=14.8\pm2.4$\,pc, which agrees with previous determinations using 
stars. As a consequence of the completeness effects, the observed radial distribution of OCs 
with respect to Galactocentric distance does not follow the expected exponential profile, 
instead it falls off both for regions external to the Solar circle and more sharply towards the 
Galactic center. We simulate the effects of completeness assuming that the observed distribution 
of the number of OCs with a given number of stars above the background, measured in a restricted zone 
outside the Solar circle, is representative of the intrinsic distribution of OCs throughout the 
Galaxy. Two simulation models are considered in which the intrinsic number of observable stars are
distributed: {\em (i)} assuming the actual positions of the OCs in the sample, and {\em (ii)} 
random selection of OC positions. As a result we derive completeness-corrected radial distributions 
which agree with exponential disks throughout the observed Galactocentric distance range 5--14\,kpc, 
with scale lengths in the range $\rm\rd=1.5 - 1.9\,kpc$, smaller than those inferred by means of stars. 
In particular we retrieve the expected exponential-disk radial profile for the highly depleted
regions internal to the Solar circle. The smaller values of \rd\ may reflect intrinsic differences in 
the spatial distributions of OCs and stars. We derive a number-density of Solar-neighbourhood (with 
distances from the Sun $\rm\ds\leq1.3\,kpc$) OCs of $\rm\po=795\pm70\,kpc^{-3}$, which implies a total 
number of (Trumpler types I to III) OCs of $\sim730$ of which $\sim47\%$ would already have been 
observed. Extrapolation of the completeness-corrected radial distributions down to the Galactic 
center indicates a total number of OCs in the range $\rm(1.8 - 3.7)\times10^5$. These estimates are  
upper-limits because they do not take into account depletion in the number of OCs by dynamical effects 
in the inner parts of the Galaxy. The observed and completeness-corrected age-distributions of the 
OCs can be fitted by a combination of two exponential-decay profiles which can be identified with the 
young and old OC populations, characterized by age scales of $\rm\sim100\,Myr$ and $\rm\sim1.9\,Gyr$, 
respectively. This rules out evolutionary scenarios based on constant star-formation and OC-disruption 
rates. Comparing the number of observed embedded clusters and candidates in the literature with the 
expected fraction of very young OCs, derived from the observed age-distribution function, we estimate 
that 3.4--8\% of the embedded clusters do actually emerge from the parent molecular clouds as OCs.

 }

\titlerunning{Spatial distribution of open clusters}

\authorrunning{C. Bonatto et al.}

\maketitle

\section{Introduction}
\label{intro}

Understanding the true shape of the Milky Way is a challenging task which depends essentially 
on the availability of as complete as possible Galactic surveys of the spatial distribution of 
stars, star clusters, \ion{H}{i} gas, molecular clouds, etc. These observations can be used to 
better constrain theoretical mass models of the Galaxy. The modern view of the Milky Way pictures 
it as composed basically of the center and nuclear bulge, a thin disk hosting most of the OCs, a 
thick disk containing the old OCs, and the extended halo hosting the old stars, white dwarfs and 
globular clusters (e.g. Majewski \cite{Maj93}). The gas and stellar density of the disk is usually 
expressed as a combination of exponential-decay profiles for the horizontal and vertical directions,
$\rm\rho(r,z)\propto\,e^{-(r/R_D)}\,e^{-(|z|/z_h)}$, where \rd\ is the scale length and \zh\ the 
scale height (e.g. Binney \& Tremaine \cite{BinTre1987}). 

The kinematics and spatial distribution of large numbers of stars have been the best used 
probes of the Galaxy structure. For instance, M dwarfs are found to distribute vertically
with relatively large scale heights, $\rm\zh\sim300\,pc$, in contrast to the younger A-type 
stars with $\rm\zh\sim100\,pc$ (Faber et al. \cite{Faber76}). Giants and subgiants have scale 
heights typically between 200 and 500\,pc, though the uncertainty is larger than that of the 
dwarf luminosity class (e.g. Bahcall \& Soneira \cite{BS81}). Using star counts from the Sloan 
Digital Sky Survey taken from two high-latitude ($\rm49^\circ<|b|<64^\circ$) samples north and 
south of the Galactic plane Chen, Stoughton, Smith et al. (\cite{ChenSS01}) derived 
$\rm\zh=330\pm3\,pc$ for the old thin disk and $\rm\zh=580-750\,pc$ for the thick disk. They 
suggest that the thick disk formed through heating of a preexisting thin disk caused by the 
merging of a satellite galaxy. Based on the kinematics of sdB stars de Boer et al. 
(\cite{BAA97}) and Altmann et al. (\cite{AEB04}) found that these stars are characterized by 
two distributions, a disk one with $\rm\zh\sim1\,kpc$ and an extended one with 
$\rm\zh\sim7\,kpc$. In recent years considerable amounts of data on parallax and proper 
motion of stars has allowed to probe in detail the Galactic structure in distances of up to 
2\,kpc from the Sun. In particular Kaempf, de Boer \& Altmann (\cite{KBA05}) using the 
kinematics and spatial distribution of red horizontal-branch stars selected from the 
HIPPARCOS catalogue derived a disk scale height of $\rm\zh\approx0.6\,kpc$.

With respect to the vertical position of the Sun (\zo), asymmetries in star counts above 
and below the Galactic plane have put it displaced a few parsecs to the North. Several 
methods used in the past constrained the value of \zo\ to the range 4--40\,pc (see, e.g. the 
account in Cohen \cite{Cohen95}). More recently, different sets of data have been used to 
obtain more accurate values of \zo. Cohen (\cite {Cohen95}), using IRAS point-source counts 
at 12 and $\rm25\,\mu\,m$ obtained $\rm\zo=15.5\pm0.7\,pc$. Hammersley et al. (\cite{HGMC95}) 
examined star counts and surface brightness maps from COBE, IRAS and the Two-micron 
Galactic Survey to derive $\rm\zo=15.5\pm3\,pc$. The high-Galactic latitude data of Chen, 
Stoughton, Smith et al. (\cite{ChenSS01}) resulted in $\rm\zo=27\pm4\,pc$.

Measurements of the horizontal scale length by means of stars place \rd\ in the range from 
2.2 to 3.5\,kpc (de Vaucouleurs \& Pence \cite{deVP78}; Knapp, Tremaine \& Gunn \cite{Knapp78}; 
McCuskey \cite{McC69}).

Open clusters are formed in, and distribute throughout the disk. Interactions with the disk 
and the tidal pull of the Galactic center/bulge tend to destroy the poorly-populated OCs in a 
time-scale of a few $\rm10^8\,Myr$ (Bergond, Leon \& Guibert \cite{Bergond2001}). The dynamical
disruption is more critical particularly for the OCs more centrally located. The survivors, 
however, may reach large vertical distances from the plane spanning the thick disk. Although 
the number of known OCs is small compared to stars, it is relatively simple and accurate to 
derive distance and age for these objects. In this sense, OCs - in particular their spatial 
distribution - may serve as a direct probe of the disk structure. By means of a sample of 
72 OCs Janes \& Phelps (\cite{JP94}) found that the young OCs ($\rm age\leq800\,Myr$) are 
distributed almost symmetrically about the Sun with a scale height perpendicular to the 
Galactic plane of $\rm\zh\approx55\,pc$, while for the old OCs they found $\rm\zh\approx375\,pc$.

It is interesting to point out that since Janes \& Adler (\cite{JA82}) study the number of 
OCs with derived parameters increased nearly 50\%, taking as reference the clusters 
currently with parameters in the WEBDA\footnote{\em http://obswww.unige.ch/webda} OC 
database (Mermilliod \cite{Merm1996}). However, in the recent revision of OCs Dias et al. 
(\cite{Dias2002}) report a total of 1537 catalogued OCs. This clearly indicates the 
need to explore more systematically the available OC databases. WEBDA gathers data from
a variety of sources, which implies different observational instrumentation and analysis 
methods and, consequently uncertainties in the values of parameters. We deal with the 
non-uniformity of the parameters by {\em (a)} working with bins wide enough to take into 
account most of the uncertainties, and {\em (b)} assuming $\rm 1\sigma$ Poisson errors to 
assign statistical significances to the results.

In this paper we build an OC sample composed of the WEBDA objects added to 10 other OCs 
studied in recent papers. We intend to use this OC sample, particularly the z-measurements, 
to derive the disk scale height (\zh), the displacement of the Sun above the Galactic plane 
(\zo), and the horizontal scale length. In addition, we use the present OC sample to study 
statistical properties related to the reddening, distance from the Sun, location in the Galaxy,
completeness and, particularly, to derive the intrinsic OC age-distribution function. 

Previous statistical works on the relation of OC parameters with Galactic structure are,
e.g. Lyng\aa\ (\cite{Lynga82}), Janes \& Phelps (\cite{JP94}), Nilakshi et al. 
(\cite{Nilakshi2002}), and Tadross et al. (\cite{Tad2002}). The main results of Lyng\aa\ 
(\cite{Lynga82}), related to the present work, are {\em (i)} a correlation of cluster age 
with Galactocentric distance, {\em (ii)} an exponential OC disintegration time scale of 
$\rm\sim10^8\,yr$, {\em (iii)} displacement of the Sun over the Galactic plane of 20\,pc, 
and {\em (iv)} a z-distribution which differs only for the oldest clusters. Nilakshi et al. 
(\cite{Nilakshi2002}) found that OCs with Galactocentric distances larger than 9.5\,kpc have 
larger sizes. Tadross et al. (\cite{Tad2002}) found {\em (i)} evidence that older clusters 
are found preferentially in the outer parts of the disk, and {\em (ii)} that most of the 
OCs younger than $\rm32\,Myr$ are concentrated on the Perseus arm. 

This paper is organized as follows. In Sect.~\ref{OCS} we present the OC sample and discuss 
statistical properties related to the spatial and age distributions. In Sect.~\ref{complet} 
we discuss qualitative aspects affecting sample completeness. In Sect.~\ref{dsh} 
we derive the displacement of the Sun above the Galactic plane and the observed scale height. 
In Sect.~\ref{SimCom} we simulate the effects of sample completeness. In  Sect.~\ref{Exten}
we derive completeness-corrected radial and vertical distribution functions, and discuss
properties of completeness-corrected age distribution functions. In Sect.~\ref{tocg} 
inferences on the total number of OCs in the Galaxy are presented. Concluding remarks are in 
Sect.~\ref{Conclu}.

\section{Statistical characterization of the OC sample}
\label{OCS}

The OCs used in the present study are those with Galactic coordinates, distance from the Sun, 
reddening and age catalogued in WEBDA. To these we add 10 recently studied OCs (2 in Bica, 
Bonatto \& Dutra \cite{BiBoDu03}; 3 in Bonatto, Bica \& Dutra \cite{BoBiDu04}; and 5 in 
Ortolani, Bica \& Barbuy \cite{OBB05}). The sample totals 654 OCs with known distance from the
Sun (of which 636 have age determination).

The significant number of OCs in this sample presents an opportunity to study properties of
clusters in different age ranges separately. We selected the age ranges {\em (i)} 
$\rm age\leq200\,Myr$ (young OC population), {\em (ii)} $\rm200\,Myr\leq age\leq1\,Gyr$
(moderate-age OCs), and {\em (iii)} $\rm age\geq1\,Gyr$ (old OCs) both for statistical 
and population representativity purposes. The number of clusters in each age
range is 402 (young), 148 (moderate-age) and 86 (old).
 
\subsection{Spatial distribution}
\label{SpaDis}

\begin{figure} 
\resizebox{\hsize}{!}{\includegraphics{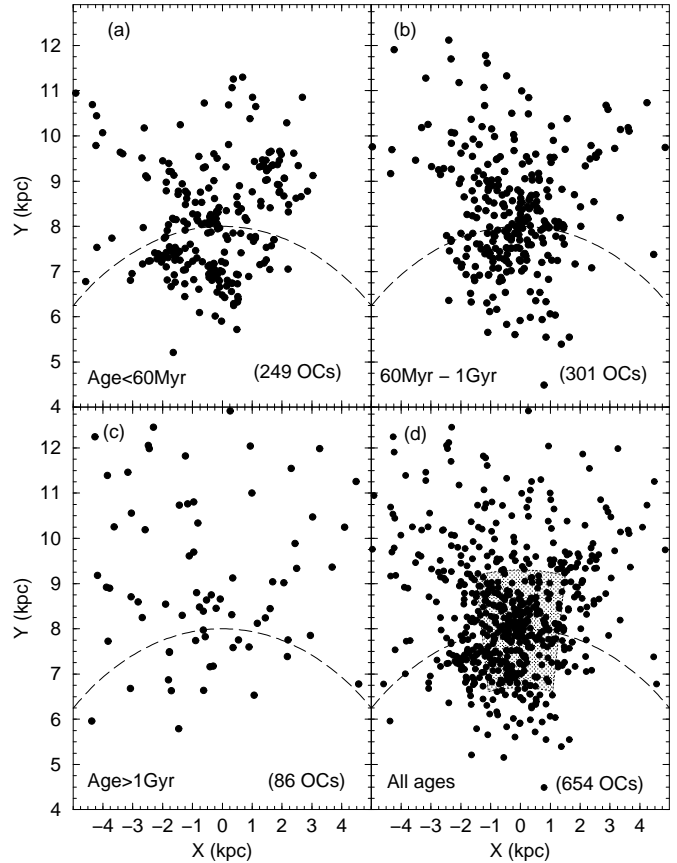}}
\caption[]{Spatial distribution of the OCs, according to age groups. The Solar circle is
shown by the dashed line, and the Galactic center is located at (0,0). Shaded area in 
panel (d): region encompassed by the restricted sample (Sect.~\ref{tocg}). For 
comparison purposes the coordinate definitions follow Janes \& Phelps (\cite{JP94}).}
\label{fig1}
\end{figure}

In Fig.~\ref{fig1} we show the spatial distribution of the OCs in the present sample projected 
onto the X and Y plane, separated according to age groups. Throughout this paper
we use 8.0\,kpc (Reid \cite{Reid93}) as the distance of the Sun to the Galactic center.
In this figure conspicuous structures of the local disk are apparent. 
For clarity we show the Solar circle in all panels. Note that in panels (a) and (b) the 
threshold between young and moderate ages has been changed to 60\,Myr in order to better
isolate OCs related to spiral arms. OCs younger than 60\,Myr (panel (a)) trace out the 
local distribution of the Orion Arm (Georgelin \& Georgelin \cite{GG70}), in which the Sun is 
located close to the inner border. Another conspicuous feature defined by the young OCs is the gap 
towards the Galactic center at $\rm\dgc\approx7.5\,kpc$ related to the dip before the Sgr-Car 
Arm. Clusters older than 1\,Gyr (panel (c)) are uniformly distributed but with an asymmetry in 
the number of OCs towards the anti-center, a region where they are well-known to populate 
(Friel \cite{Friel95}). The all-ages sample is shown in panel (d) where traces of the Orion 
Arm, the gap at $\rm\dgc\approx7.5\,kpc$ and the radial asymmetry can still be detected.

\subsection{Observed age histogram}

\begin{figure} 
\resizebox{\hsize}{!}{\includegraphics{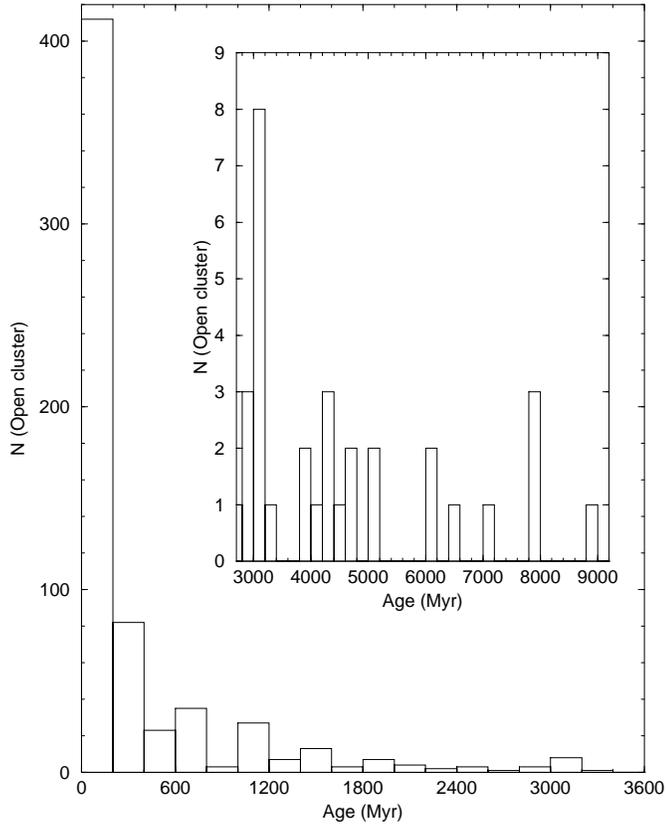}}
\caption[]{Age histogram of the OC sample for $\rm age\leq3800\,Myr$. Peaks occur at 
$\rm\sim100\,Myr$, $\rm\sim700\,Myr$, and $\rm\sim3.1\,Gyr$. The inset shows the
histogram for $\rm age\geq2800\,Myr$.}
\label{fig2}
\end{figure}

The observed age histogram of the present OCs is shown in Fig.~\ref{fig2} with bins of
200\,Myr which are wide enough to accommodate most of the uncertainties related to the
different age-determination techniques included in the WEBDA sample. Two conspicuous 
peaks occur at $\rm\sim100\,Myr$ (characterizing the young OCs) and $\rm\sim700\,Myr$
(moderate-age OCs). A less-pronounced peak occurs at $\rm\sim3.1\,Gyr$, which can be 
associated to the old OC distribution.

The age-distribution function will be presented and discussed in Sect.~\ref{tau_CC}.

\begin{figure} 
\resizebox{\hsize}{!}{\includegraphics{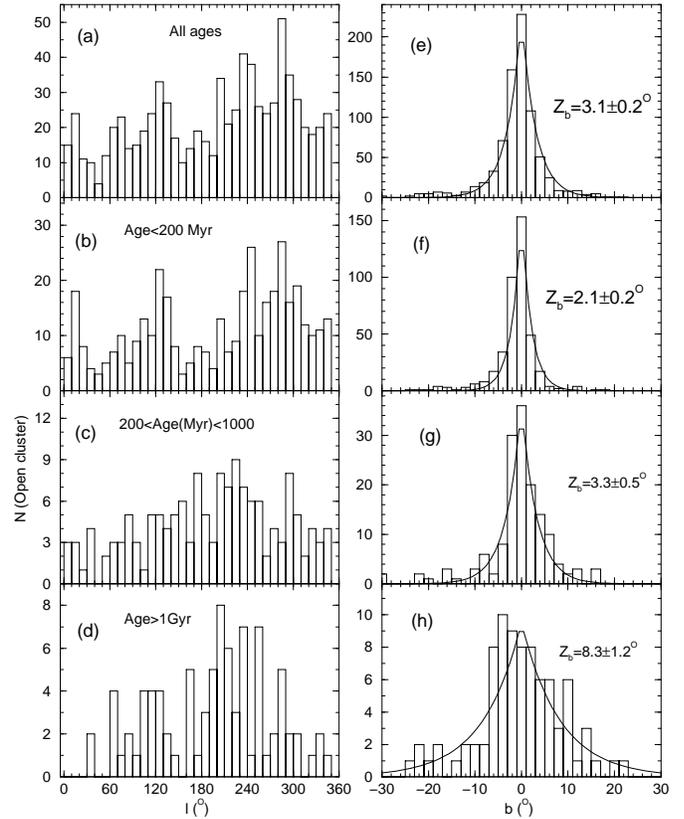}}
\caption[]{Distribution of OCs in terms of Galactic longitude (left panels) and
latitude (right panels). Solid lines: least-squares fit with the function
$\rm N\propto e^{-|b/z_b|}$. Panels (a) and (e): OCs of all ages. Panels (b) and
(f): OCs younger than 200\,Myr. Panels (c) and (g): OCs with age in the range 
200\,Myr -- 1\,Gyr. Panels (d) and (h): OCs older than 1\,Gyr.}
\label{fig3}
\end{figure}

\subsection{Galactic coordinates}

In the left panels of Fig.~\ref{fig3} we examine the distribution of OCs in terms of 
Galactic longitude. In all age ranges the number of OCs drops off towards the Galactic 
center and increases roughly towards the third quadrant. These facts might reflect both 
an observational limitation in the sense that low-contrast clusters are more difficult to 
detect in directions projected against the Galaxy center/bulge (or more simply, observers 
tended to avoid crowded central fields -- Sect.~\ref{complet}) and the enhanced tidal-disruption 
probability of OCs closer to the Galactic center/bulge (Bergond, Leon \& Guibert \cite{Bergond2001}).

The distribution of OCs in terms of Galactic latitude is in the right panels of Fig.~\ref{fig3}. 
As expected, most of the OCs are tightly concentrated close to the Galactic plane (panel (e)). 
However, the distributions become wider for older age groups (panels (f) to (h)). The
distributions can be analytically described by the 
exponential-decay function $\rm N\propto e^{-|b/z_b|}$. We find $\rm z_b=2.10\pm0.20^\circ$ 
for the young OCs (panel (f)), $\rm z_b=3.47\pm0.51^\circ$ for the moderate-age OCs 
(panel (g)), and $\rm z_b=8.27\pm1.13^\circ$ for the old OCs (panel (h)). The all-ages
distribution (panel (e)) is characterized by $\rm z_b=3.14\pm0.21^\circ$. The 
asymmetry observed in the old-age distribution with respect to Galactic latitude (panel (h))
may be accounted for by completeness (Sects.~\ref{complet} and \ref{SimCom}). Because these 
clusters are not as concentrated near the Galactic plane as the young OCs (Fig.~\ref{fig7}),
the old OCs located above the Galactic plane become easier to detect than the ones in
the opposite side. A similar, but less-conspicuous effect can be seen in the distribution 
of the moderate-age OCs (panel (g)). Alternatively, this asymmetry might reflect an intrinsic
difference in the distribution of old OCs above and below the plane.

\subsection{Distance from the Sun and reddening}

\begin{figure} 
\resizebox{\hsize}{!}{\includegraphics{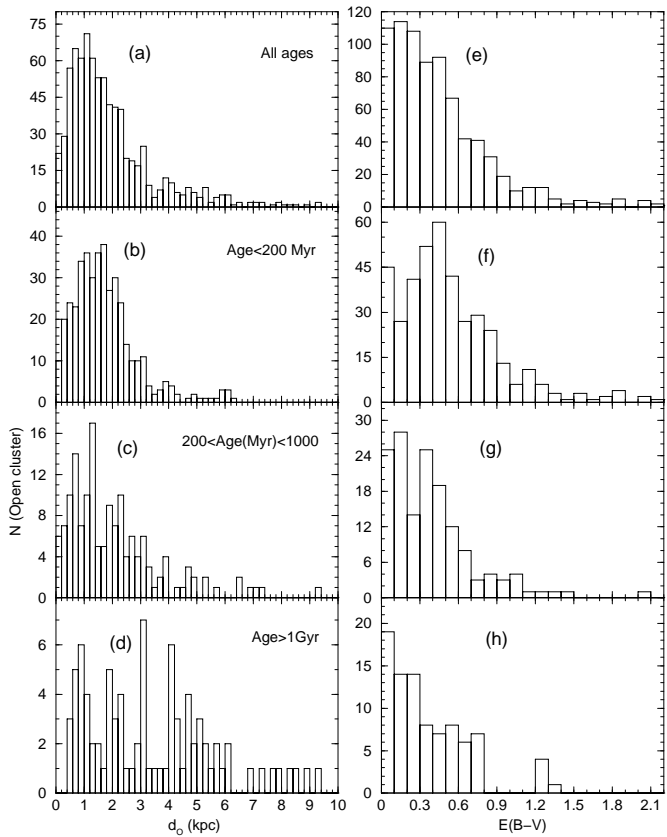}}
\caption[]{Distribution of OCs in terms of distance from the Sun (left panels)
and reddening (right panels). Panel ages as in Fig.~\ref{fig3}.}
\label{fig4}
\end{figure}

In the left panels of Fig.~\ref{fig4} we show the distribution of OCs with respect to the 
distance from the Sun, in bins of 0.2\,kpc. The young (panel (b)) and moderate-age 
(panel (c)) OCs present similar distributions with peaks at $\rm\ds\approx1.7\,kpc$
and $\rm\ds\approx1.3\,kpc$, respectively. 
Observational limitations probably affect the detection of low-contrast OCs for 
$\rm\ds\ge1.7\,kpc$. The old OCs (panel (d)) are more evenly distributed 
(Fig.1, panel (d)).

\begin{figure} 
\resizebox{\hsize}{!}{\includegraphics{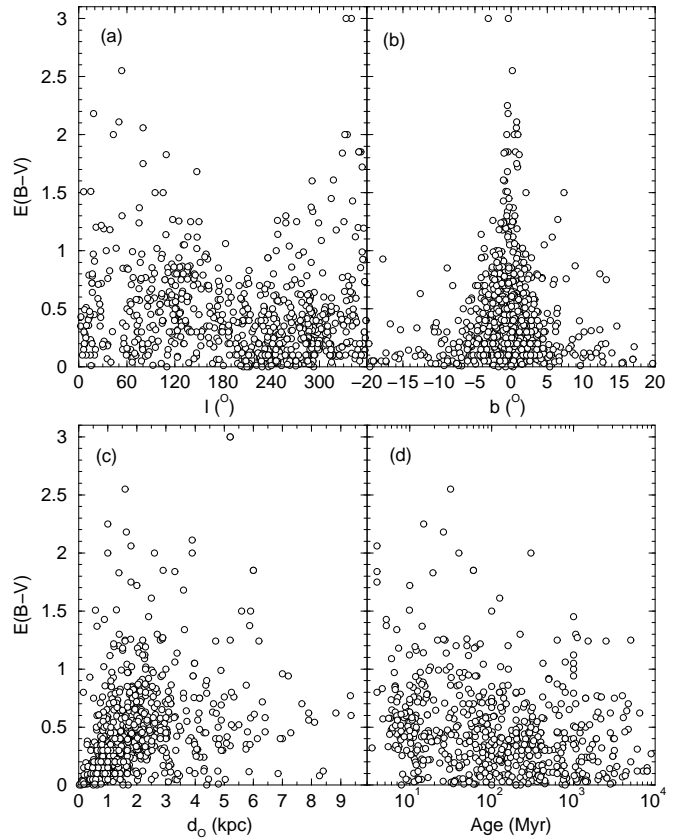}}
\caption[]{Dependence of reddening on Galactic longitude (panel (a)), latitude (panel (b)),
distance from the Sun (panel (c)) and cluster age (panel (d)). Each point represents an OC.}
\label{fig5}
\end{figure}

With respect to the reddening (right panels of Fig.~\ref{fig4}) the distribution of the 
young OCs (panel (f)) is wider than those of the moderate-age (panel (g)) and old 
(panel (h)) OCs. This fact probably can be accounted for by more significant internal 
reddening affecting young OCs. In addition, excess interstellar reddening also affects
more significantly young OCs, since these clusters are more tightly concentrated to
the Galactic plane (panel (f) in Fig.~\ref{fig3}). The strong dependence of reddening on
OC location in the Galaxy can be appreciated in panels (a) and (b) of Fig.~\ref{fig5}.
As expected, the reddening increases markedly towards the Galactic center/bulge (1st and 4th
quadrants in panel (a)) and plane (the conspicuous peak in panel (b)). \ebv\
increases nearly linearly with the distance from the Sun up to $\rm\approx3\,kpc$
(panel (c)); the scatter increases significantly for larger distances. The excess
reddening observed in the young OCs is apparent as well in panel (d).

\subsection{Galactocentric distance}
\label{GalDis}

\begin{figure} 
\resizebox{\hsize}{!}{\includegraphics{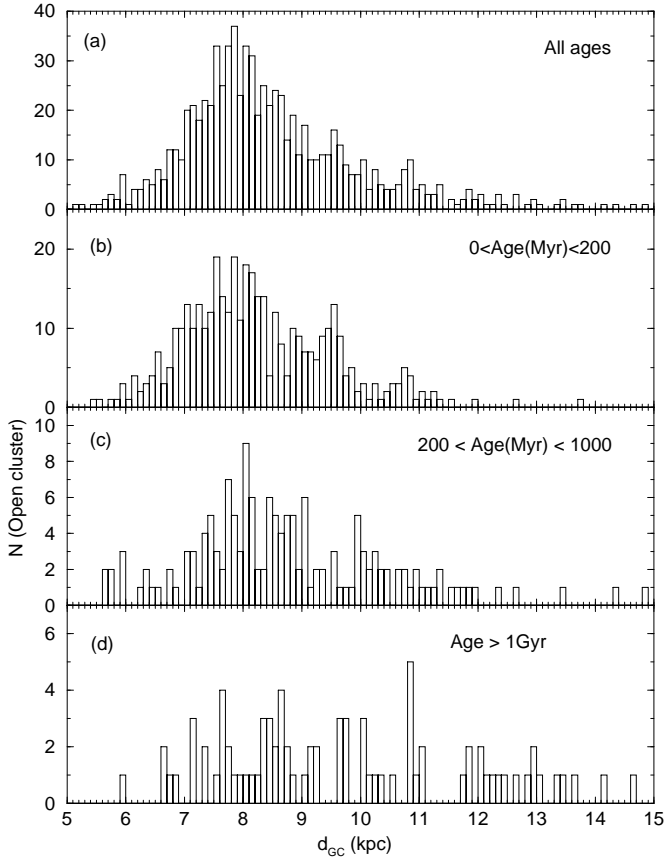}}
\caption[]{Distribution of OCs in terms of Galactocentric distance. Panel ages as in 
Fig.~\ref{fig3}.}
\label{fig6}
\end{figure}

In Fig.~\ref{fig6} we examine histograms with the number of observed OCs in bins of 
Galactocentric distance. Similarly to the distance from the Sun, the old OCs (panel (d)) are 
more evenly distributed than the young (panel (b)) and moderate-age (panel (c)) OCs. The radial 
distribution of the all-ages OCs (panel (a)) is not symmetrical about the Galactocentric 
distance of the Sun, instead it presents a maximum at the Solar position and drops off more 
sharply towards the Galactic center than outside the Solar circle. OCs younger than
200\,Myr and those with age in the range 200\,Myr -- 1\,Gyr present similar radial
distributions. The observed radial distribution of the OCs clearly does not resemble 
that expected of an exponential disk (e.g. Binney \& Tremaine \cite{BinTre1987}). We will 
return to this point in Sect.~\ref{tocg}.

\section{Qualitative considerations on sample completeness}
\label{complet}

In general terms, cluster detection-rates depend on factors such as the stellar content (both 
in absolute numbers and the presence of giants), apparent magnitude at the turnoff, distance
from the Sun, stellar background projected onto the OC direction, etc. External factors such
as systematic surveys with deep photometry are necessary as well to detect as many OCs as 
possible. In addition, the identification of a stellar concentration as a star cluster depends 
on the analysis of the colour-magnitude diagram morphology, the structure of radial-density 
profiles, and proper-motion data, when available (e.g. Bonatto \& Bica \cite{BB2005}; Bica \& 
Bonatto \cite{BiBo2005a}; Bica \& Bonatto \cite{BiBo2005b}, and references therein). 

The present sample is not complete, particularly with respect to OCs at large distances from the 
Sun and/or low-contrast clusters. The cluster data in WEBDA come from photographic, photoelectric
or CCD photometry taken with different instrumentation over the last 50 years. This heterogeneity 
of data sources precludes objective inferences on the faint magnitude-limit of the stars, total 
number of stars, surface brightness or density contrast in the present OCs. Basically the WEBDA 
sample contains OCs of the denser Trumpler types I to III (Ruprecht \cite{Rup66}), because observers 
naturally directed their attention to this kind of object. 

The following argument illustrates the difficulties associated with the identification of low-contrast 
clusters in different regions of the Galaxy. Assume an OC whose radial density profile can 
be described by the two-parameter King (\cite{King1966}) law, $\rm\sigma(r)=\sigma_{bg}+
\frac{\sigma_{0K}}{1+(r/R_c)^2}$, where $\rm\sigma_{bg}$ is the background surface density of 
stars, $\rm\sigma_{0K}$ is the central density of stars (above the background), and $\rm R_c$ is 
the core radius. In fact, the background represents the contamination of non-cluster stars projected
onto the OC angular area, integrated along the direction towards the OC, fore and background stars alike. 
In this sense the background density of an OC is a function of $\ell$ and $b$ only, and not of
distance from the Sun (Sect.~\ref{SimCom}).

We define the contrast parameter as 
$\rm\delta_c=\frac{\sigma(0)}{\sigma_{bg}}=1+\frac{\sigma_{0K}}{\sigma_{bg}}$. Integration of 
$\rm\sigma(r)$ from the center to the limiting radius ($\rm R_{lim}$) and considering that
$\rm R_{lim}\geq5\times R_c$ (Bonatto \& Bica \cite{BB2005}), the total number of cluster stars 
(background-subtracted) is $\rm N*\approx2\pi\sigma_{0K} R_c^2\ln(R_{lim}/R_c)$. Thus, the 
contrast parameter as a function of the background density and effective number of stars can 
be expressed as 

\begin{equation}
\label{eq_delta}
\rm\delta_c\approx1+\frac{\left(N*/\sigma_{bg}\right)}{2\pi\ln(R_{lim}/R_c)R_c^2}.
\end{equation} 

Extractions of 2MASS\footnote {The Two Micron All Sky Survey, All Sky data release (Skrutskie 
et al. \cite{2mass1997}), {\em http://www.ipac.caltech.edu/2mass/releases/allsky/}} photometry 
in directions which do not intercept detected clusters show that the average background stellar 
density in low-latitude ($|b|\leq5^\circ$) regions towards the Galactic center/bulge is a factor 
$\rm\sim10$ as large as that towards the anti-center (Sect.~\ref{SimCom}). Thus, an anti-center 
cluster with $\rm\delta_c\sim11$ would have $\rm\delta_c\sim2$ when projected against the center/bulge. 
In this sense, if the typical detection limit of WEBDA OCs is restricted to, say $\rm\delta_c\ge 2$, 
surveys towards the Galactic center/bulge would miss the whole class of OCs with $\rm\delta_c$ in 
the range 2 to 11 which in principle would be detected towards the anti-center. 

To give an idea of the values $\rm\delta_c$ takes on with actual OCs we quote here some examples 
taken from Bonatto \& Bica (\cite{BB2005}) and Bica \& Bonatto (\cite{BiBo2005b}), derived with 
2MASS data. The high-Galactic latitude OCs NGC\,188 ($b=+22.39^\circ$; $\ds\sim1.7$\,kpc; 
$\rm M\sim3800\,\ms$) and M\,67 ($b=+31.89^\circ$; $\ds\sim0.9$\,kpc; $\rm M\sim990\,\ms$) present 
$\rm\delta_c=22.5\pm2.9$ and $\rm\delta_c=34.7\pm5.7$, respectively, while the populous and rather 
low-latitude cluster NGC\,2477 ($b=-5.82^\circ$; $\ds\sim1.2$\,kpc; $\rm M\sim5300\,\ms$) has 
$\rm\delta_c=20.7\pm1.9$. On the other hand, the relatively low-contrast, third-quadrant OCs 
Ruprecht\,78 ($b=-1.88^\circ$; $\ds\sim1.3$\,kpc; $\rm M\sim860\,\ms$), Haffner\,4 ($b=-3.62^\circ$; 
$\ds\sim1.7$\,kpc; $\rm M\sim260\,\ms$) and Trumpler\,13 ($b=-2.34^\circ$; $\ds\sim1.9$\,kpc; 
$\rm M\sim420\,\ms$), present $\rm\delta_c=4.2\pm2.6$, $\rm\delta_c=4.3\pm2.2$, and 
$\rm\delta_c=8.9\pm5.5$, respectively.

Completeness is as well affected by cluster distance from the Sun in depth-limited photometry. 
OC stars fainter than some apparent magnitude may present exceedingly large photometric errors 
and consequently, end up not being separated from the background. To estimate the dependence of 
this effect on distance from the Sun we consider an artificial OC characterized by a standard 
Salpeter (Salpeter \cite{Salp55}) mass function ($\rm\phi(m)\propto m^{-(1+\chi)}$, with 
$\rm\chi=1.35$) with the turnoff at $\rm m=5\,\ms$. Based on our previous experience in the analysis 
of OCs with 2MASS data (e.g. Bonatto \& Bica \cite{BB2005}, and references therein) we take stars 
with $\rm J=15$ as representative of the average faint-magnitude limit which still can be used to 
derive OC structural parameters. Using the mass-luminosity relation derived from the 10\,Myr, 
Solar-metallicity Padova isochrone (Girardi et al. \cite{Girardi2002}), the low-mass limit 
corresponding to $\rm J=15$ can be expressed as a function of distance from the Sun as 
$\rm m_{low}(\ds)=-1.18+0.6\log(\ds)+3.4\times10^{-6}\ds^{3/2}$. Integration of the mass function 
from the low-mass limit to the turnoff yields an estimate of the effective number of stars above 
the background, $\rm N*$. To a good approximation, this number can be expressed as a function of 
distance from the Sun (for $\rm0.2\leq\ds(kpc)\leq6$) as $\rm N*(\ds)\propto(\ds/1kpc)^{-1}$. Thus, 
a given OC placed twice as distant from the Sun ends up presenting only about half of the stars above 
the background. Roughly speaking, this effect is similar to doubling the value of $\sigma_{bg}$ in 
eq.~\ref{eq_delta}, which in turn produces a smaller value of $\rm\delta_c$. 

From the observational point of view, recent proper-motion and parallax data have been used to 
explore the low-contrast end (even less concentrated than Trumpler type IV) of the OC distribution. 
An example is the use of HIPPARCOS data to search for poorly-populated OCs or candidates in the 
solar-vicinity ($\rm\ds\leq500\,pc$) by Platais, Kozhurina-Platais \& van Leeuwen (\cite{Platais98}) 
which yielded about 10 such objects which have large angular sizes. One must be aware of the 
completeness effects discussed above, but it is also interesting to point out that at least for the 
Solar-vicinity the number of extremely-low contrast OCs appears not to be exceedingly large. 

We conclude from the above that the present sample (and the results derived therefrom) -  although
certainly affected by completeness - can be taken as statistically representative of OCs of Trumpler 
types I to III.

The discussions above raise important points related to sample completeness and make it clear 
the need to undertake efforts to survey more thoroughly with deeper photometry - at least the 
Solar neighbourhood - to complete the local census of the OCs, particularly the low-contrast ones.
However, they alone do not allow us to objectively quantify the completeness either of a particular 
OC or the average completeness of the whole sample. We will return to this issue in Sect.~\ref{SimCom} 
where the quantitative effects of completeness will be derived by means of simulation.

\section{The observed disk scale height}
\label{dsh}

\begin{figure} 
\resizebox{\hsize}{!}{\includegraphics{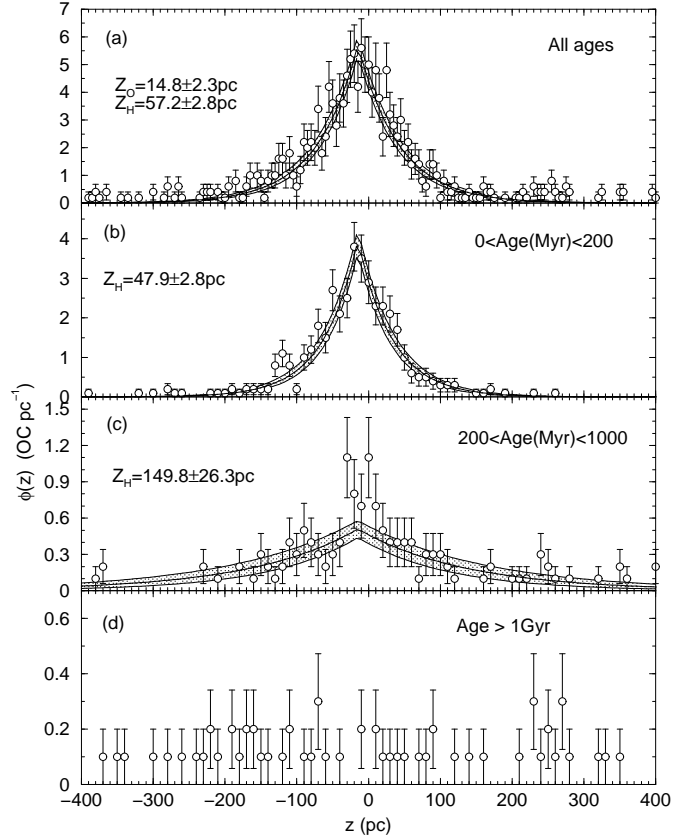}}
\caption[]{Distribution of OCs in terms of height over the Galactic plane. Age groups
are considered separately. Solid lines: least-squares fit with 
$\rm\phi(z)=\phi_0e^{-|(z+z_\odot)/z_h|}$. Shaded areas: $\rm1\sigma$-standard
deviation. Top panel: bins of 5\,pc in z. Bottom panels: 10\,pc bins.}
\label{fig7}
\end{figure}

We calculated the vertical position z of all OCs with Galactic longitude and 
latitude, and distance from the Sun. The resulting distributions are shown in
Fig.~\ref{fig7} for the different age groups. The uncertainties correspond to 
$\rm 1\sigma$ Poisson errors. Similarly to the Galactic latitude (Fig.~\ref{fig3}), 
the z-distributions also depend on OC age, in the sense that older OCs are more 
evenly distributed in z than the young OCs. The young (panel (b)) and moderate-age 
(panel (c)) OCs have distributions following exponential-decay profiles, while the 
old OCs (panel (d)) are almost uniformly distributed in height. The total
number of OCs in each distribution is 402 (young), 148 (moderate-age) 
and 86 (old). Note that the number of clusters with $\rm \mid z\mid\geq400\,pc$, 
which are not shown in Fig.~\ref{fig7}, is 0 for the young, 12 for the moderate-age 
and 27 for the old OCs.

Thus, the near uniformity in the z-distribution of the old OCs
cannot be accounted for by small-number statistics alone, but it appears to reflect
their intrinsic distribution. 

In order to reach a higher z-resolution we built the all-ages distribution (panel (a))
with bins of 5\,pc, while for the remaining distributions the z-bin is 10\,pc. The higher
z-resolution is necessary in order to derive more accurately the distance of the Sun from
the Galactic plane while at the same time keeping the $\rm 1\sigma$ Poisson errors
non-prohibitively large.

We fit the all-ages distribution in panel (a) with the function $\rm\phi(z)=\phi_0e^{-|(z+z_\odot)/z_h|}$,
where \zo\ is the displacement of the Sun with respect to the Galactic plane and \zh\ is the vertical 
scale height. For this particular analysis we center the coordinate system on the Sun. We obtain 
$\rm\zo=14.8\pm2.3\,pc$ (above the plane), which within the uncertainties, is in excellent agreement 
with the values calculated by Cohen (\cite {Cohen95}) and Hammersley et al. (\cite{HGMC95}) using stars. 
This value of \zo\ was kept fixed when fitting the young and moderate-age OCs distributions. The results 
are in Table~\ref{tab1}. It is remarkable that large samples of different objects (stars and OCs)
provide essentially the same value of the Solar displacement from the plane. For completeness
we also applied the unconstrained exponential-decay fit to the young and moderate-age OCs. For the
young OCs we obtain $\rm\zo=14.2\pm2.3\,pc$  and $\rm\zh=47.9\pm2.8\,pc$ which, within the
uncertainties are in excellent agreement with the constrained values (Table~\ref{tab1}). However, the
unconstrained fit of the less-uniform distribution of the moderate-age OCs produced a large uncertainty 
in \zo, $\rm9.8\pm11.4\,pc$ and a similar value of \zh, $\rm145\pm25\,pc$. We conclude that the 
value of the displacement of the Sun over the Galactic plane derived from the distribution of the young
OCs is, within the uncertainties, the same as the average value which considers OCs of all ages.
 
\begin{table}
\caption[]{Observed disk-structural parameters}
\label{tab1}
\renewcommand{\tabcolsep}{2.0mm}
\renewcommand{\arraystretch}{1.3}
\begin{tabular}{ccccc}
\hline\hline
Age range&$\rm\phi_0$ & $z_\odot$& \zh &CC \\
(Myr)&($\rm pc^{-1}$)& (pc) & (pc) & \\
(1)&(2)&(3)&(4)&(5)\\
\hline
All ages&$5.7\pm0.3$ & $14.8\pm2.4$ & $57.2\pm2.8$ & 0.87\\
$\leq200$& $4.0\pm0.3$ & $14.8\pm2.4^\dag$ & $47.9\pm2.8$ & 0.93\\
$200 - 1000$& $0.5\pm0.1$ & $14.8\pm2.4^\dag$ & $149.8\pm26.3$ & 0.64\\
\hline\hline
\end{tabular}
\begin{list}{Table Notes.}
\item $\rm\phi(z)=\phi_0\,e^{-|(z+z_\odot)/z_h|}$. Col.5: correlation coefficient.
$(^\dag)$: parameter kept fixed.
\end{list}
\end{table}

According to the above, the young OCs are distributed close to the disk with a scale height 
$\rm\zh\approx48\,pc$. The scale height derived from the moderate-age OCs is $\sim3$ times 
larger than that of the young OCs. Our value of \zh\ for the young OCs agrees with that derived 
by Janes \& Phelps (\cite{JP94}) for the similar age group. On the other hand, in the present
case the old ($\rm age\geq1\,Gyr$) clusters are uniformly distributed in z, in contrast to the 
$\rm\zh\approx375\,pc$ distribution derived by Janes \& Phelps (\cite{JP94}).

\subsection{Dependence of \zh\ on Galactocentric distance}
\label{DepenZH}

\begin{figure} 
\resizebox{\hsize}{!}{\includegraphics{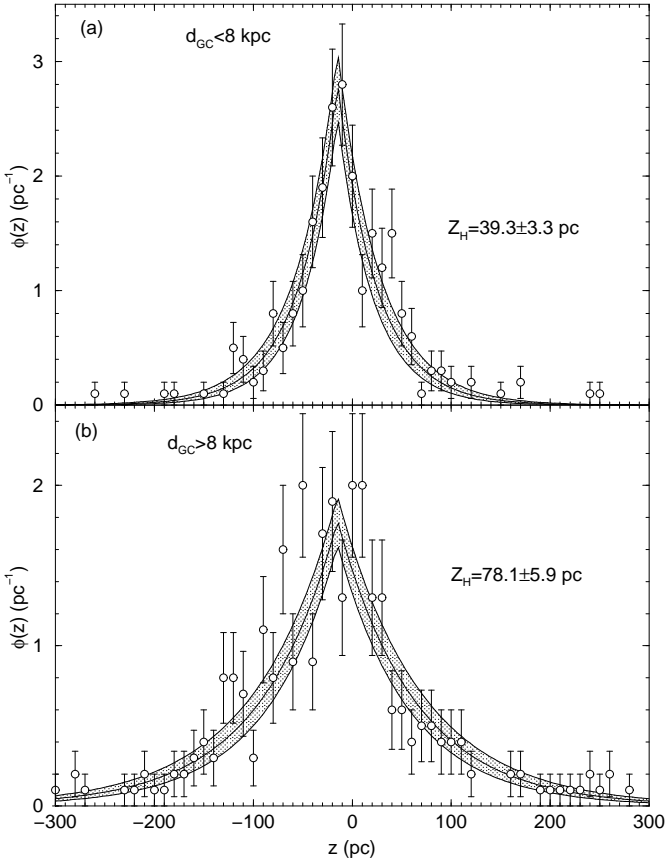}}
\caption[]{z-distribution function of OCs (younger than 1\,Gyr) more internal (panel (a)) 
and more external (panel (b)) than the Solar circle.}
\label{fig8}
\end{figure}

In recent years observational evidence in favour of disk thickening with increasing Galactocentric 
distance has been presented (e.g. Kent, Dame \& Fazio \cite{KDF91}, Janes \& Phelps \cite{JP94}). 
To examine this issue we build z-distribution functions for clusters more internal and more external 
than the Solar circle. We exclude from this analysis the old OCs because of their nearly-uniform
z-distribution (panel (d) in Fig.~\ref{fig7}). The results are in Fig.~\ref{fig8}.

From the fits we conclude that the average scale height for disk-regions outside the Solar 
circle is $\rm\zh=78.1\pm5.9\,pc$ $\rm(correlation\ coefficient\ CC=0.87$), while for regions 
interior to the Solar circle $\rm\zh=39.3\pm3.3\,pc$ ($\rm CC=0.94$). This represents an increase 
in the average $\zh$ by a factor of $\approx2$ along the observed Galactocentric distance range.

\section{Sample completeness simulation}
\label{SimCom}

As already suggested by the histograms in Fig.~\ref{fig6}, completeness effects are most likely 
affecting critically the observed radial distribution of OCs ($\rm\phi(r)$) for large distances 
from the Sun, especially towards the Galactic center/bulge. The discussions in Sect.~\ref{complet}
pointed {\em (i)} the critical r\^ole played by the background stellar density in the detection 
of low-contrast OCs, particularly in directions intercepting the Galactic center/bulge, and
{\em (ii)} the sharp decrease of observable stars above the background with increasing distance 
from the Sun. Consequently, the dependence of completeness on Galactocentric distance (as well as 
on height over the plane) has to be taken into account so that more objective results on the 
intrinsic spatial distribution of OCs can be drawn.

To minimize the effects of distance from the Sun and background density we restricted the following 
analysis to the OCs located in the shaded region shown in panel (d) of Fig.~\ref{fig1}. This region 
contains the OCs with Galactocentric distance in the range 6.7 to 9.3\,kpc which on average
are 1.3\,kpc distant from the Sun. Considering this distance along the Solar circle in both
directions the restricted region is swept out by azimuthal angles in the range $\rm-9.23^\circ$ 
to $\rm+9.23^\circ$ with respect to the Galactic center. As a consequence of the resulting geometry 
of the region, $\rm90\%$ of the OCs included in the restricted sample have distances from the Sun of 
at most $\rm\ds=1.3\,kpc$, and only $\rm3\%$ have \ds\ in the range 1.5 to 1.8\,kpc.

\begin{figure} 
\resizebox{\hsize}{!}{\includegraphics{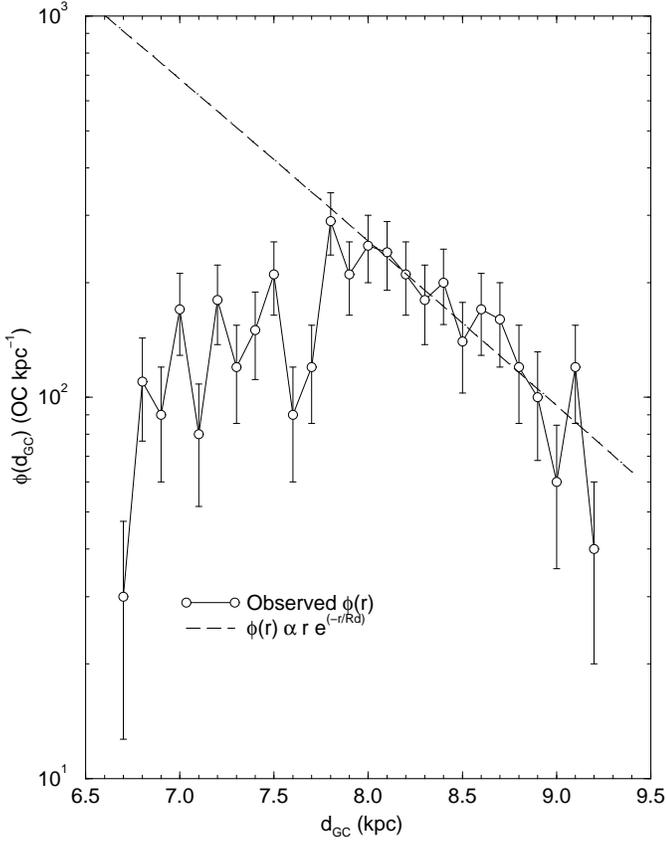}}
\caption[]{Observed distribution function of OCs in terms of Galactocentric distance in the 
restricted zone. Dashed line: exponential-disk distribution $\rm\phi(r)\propto r\,e^{-(r/R_D)}$. 
Note the depletion in the number of OCs for $\rm\dgc\leq7.8\,kpc$, and the V-shaped dip at 
$\rm\dgc\approx7.5 - 7.8\,kpc$.}
\label{fig9}
\end{figure}

In Fig.~\ref{fig9} we show the distribution function ($\rm\phi(r)=\frac{dN_{oc}}{dr}$) of OCs (in 
the restricted sample) in terms of Galactocentric distance. As already noted in Fig.~\ref{fig6} the 
observed OC distribution presents a maximum at the Sun's position and falls off both for regions 
internal and external to the Solar circle, with a V-shaped dip at $\rm\dgc\approx7.5-7.8\,kpc$ (the 
region before the Sgr-Car Arm). van den Bergh \& McClure (\cite{vdBM80}) attributed the sharp drop 
in the number of OCs for regions internal to the Solar circle to enhanced OC-disruption through the 
frequent collisions with molecular clouds in the inner regions of the Galaxy, as well as dissolution 
through tidal effects due to the Galaxy. However, when compared to the expected distribution of an 
exponential disk, $\rm\phi(r)\propto r\,e^{-(r/R_D)}$ (Sect.~\ref{tocg}), the drop in the observed 
distribution for $\rm\dgc\leq8\,kpc$ turns out exceedingly large to be accounted for by disruption 
related to dynamical effects alone. 

Considering the above facts we derive the spatial dependence of completeness by incorporating the 
qualitative arguments presented in Sect.~\ref{complet} to measurements of the background 
stellar density as seen by 2MASS photometry in different directions around the Sun. 

\begin{figure} 
\resizebox{\hsize}{!}{\includegraphics{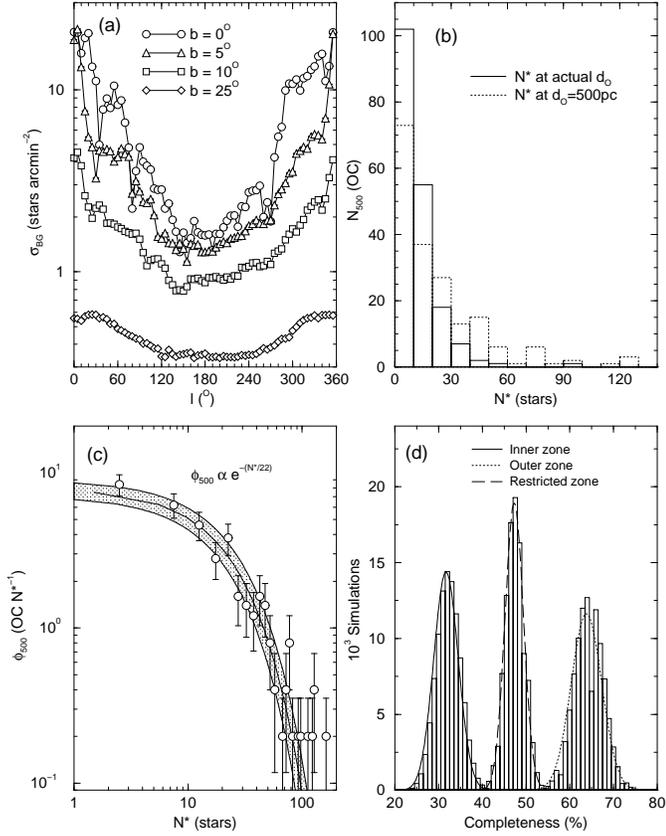}}
\caption[]{Panel (a): average background density (2MASS stars brighter than $\rm J=15$) to the north
of the Galactic plane. $\rm1\sigma$-Poisson errors are smaller than the symbols. Panel (b): histograms 
of the number of OCs (outside the Solar circle in the restricted zone) with a given effective number of 
stars at the actual distance from the Sun (solid line) and at $\rm\ds=500\,pc$ (dotted line). Panel (c): 
the distribution function at $\rm\ds=500\,pc$ can be fitted by $\rm\phi_{500}(N*)\propto e^{-(N*/22)}$; 
Shaded area: $\rm1\sigma$ standard deviation from the fit. Panel (d): average integrated completeness 
for the OCs inside (solid line) and outside (dotted line) the Solar circle and throughout the restricted 
zone (dashed line).}
\label{fig10}
\end{figure}

As a first step we measured the background stellar density of the Galaxy as a function of $\ell$ and 
$b$ using 2MASS photometry. We counted the number of stars brighter than $\rm J=15$ contained in 
circular areas with $\rm1^\circ$ in radius for the Galactic latitudes $\rm b=0^\circ, \pm2.5^\circ, 
\pm5^\circ, \pm10^\circ, \pm25^\circ\ and \pm50^\circ$ in steps of $\rm5^\circ$ in $\ell$. The 
selected circular area is large enough to accommodate most of the local fluctuations in the number 
of stars. We illustrate some of the northern background density profiles in panel (a) of Fig.~\ref{fig10}, 
where the marked difference in the average background in directions towards the Galactic center/bulge 
with respect to those towards the anti-center is clear. The corresponding southern profiles are very 
similar to the northern ones. The background contribution for a given OC (i.e. at a specific pair of 
the coordinates $\ell$ and $b$) is obtained by interpolation among the above curves.

Our analysis is based on the detectability of OCs - particularly the more distant ones - projected 
towards different regions of the Galaxy, a process which depends essentially on the density contrast 
of the central region of a cluster with respect to the field. In this sense we assume a typical angular 
radius of 1\arcmin\ to characterize the central region of the clusters. 
To estimate the effective number of stars of the OCs in the restricted zone we counted the 
number of stars brighter than $\rm J=15$ ($\rm\sim\sigma(0)$) within the $\rm R=1\arcmin$ central 
region, regardless of the OC distance from the Sun. The corresponding background contribution 
is subtracted therefrom to yield the effective number of stars, $\rm N*=\sigma(0)-\sigma_{bg}=\sigma_{0K}$. 

To avoid the high background density towards the Galactic center/bulge we restricted this procedure 
to the OCs external to the Solar circle. We assumed the WEBDA coordinates as the center of the OCs 
from which we extracted the photometry. The resulting histogram of the number of OCs with a given 
effective number of stars is shown in panel (b) of Fig.~\ref{fig10} (solid line). 
For a more objective comparison of the effective stellar content among the OCs we recalculated 
$\rm N*$ assuming $\rm\ds=500\,pc$ for all OCs, using the relation $\rm N*(\ds)\sim\ds^{-1}$ 
(Sect.~\ref{complet}). The resulting histogram is in panel (b) of Fig.~\ref{fig10} (dotted line).
Interestingly, the corresponding distribution function (panel (c)) is well represented by an 
exponential-decay profile, $\rm\phi_{500}(N*)\propto e^{-(N*/22)}$.

OCs with off-center coordinates in WEBDA and those with large projected areas such as the
Hyades and Pleiades would simply enhance the low-N* tail of the $\rm\phi_{500}(N*)$ distribution 
(panel (c) of Fig.~\ref{fig10}), which would not change appreciably the shape of the distribution.

Our basic assumption is that the observed distribution of the number of OCs with a given effective 
number of stars outside the Solar circle ($\rm\phi_{500}(N*)$) both represents the intrinsic 
distribution of OCs in the Galaxy and is essentially isotropic with respect to the Sun. Based 
on this we randomly assign to each observed OC position ($\ell$, $b$, \ds) in the restricted zone 
a value of $\rm N*$ (corresponding to $\rm\ds=500\,pc$) with a number-frequency distributed according 
to the observed histogram (panel (b) of Fig.~\ref{fig10}), and recalculate $\rm N*$ for the actual 
distance from the Sun of the OC. Finally, we divide this last number by the corresponding number of 
background stars (calculated at the respective $\ell$, $b$) to obtain the contrast parameter 
$\rm\delta_c=1+\frac{N*}{\sigma_{bg}}$\footnote{$\rm \frac{N*}{\sigma_{bg}}=\frac{\sigma_{0K}}
{\sigma_{bg}}=\delta_c-1$}. We count as detectable only the OCs with $\rm\delta_c\geq2$ (i.e. 
$\rm \frac{N*}{\sigma_{bg}}\geq1$). Completeness at a given position is then the fraction of 
retrieved OCs with respect to the input ones. To test convergence of this procedure we ran 
$\rm10^5$ simulations. We illustrate the convergence of this process in panel (d) of Fig.~\ref{fig10} 
where we plot histograms of the resulting integrated completeness value with respect to the number of 
simulations, both for the OCs inside and outside the Solar circle. The histograms are well 
represented by gaussian distributions centered at 31.7\% and 63.8\%, with standard deviations of 
2.8\% and 3.5\%, respectively for the OCs inside and outside the Solar circle. As expected because 
of the higher background density, the average completeness for the OCs inside the Solar circle is 
about half of that for the OCs outside it. The average integrated completeness throughout the entire 
restricted zone is $\rm47.3\pm2.1\%$.

\begin{figure} 
\resizebox{\hsize}{!}{\includegraphics{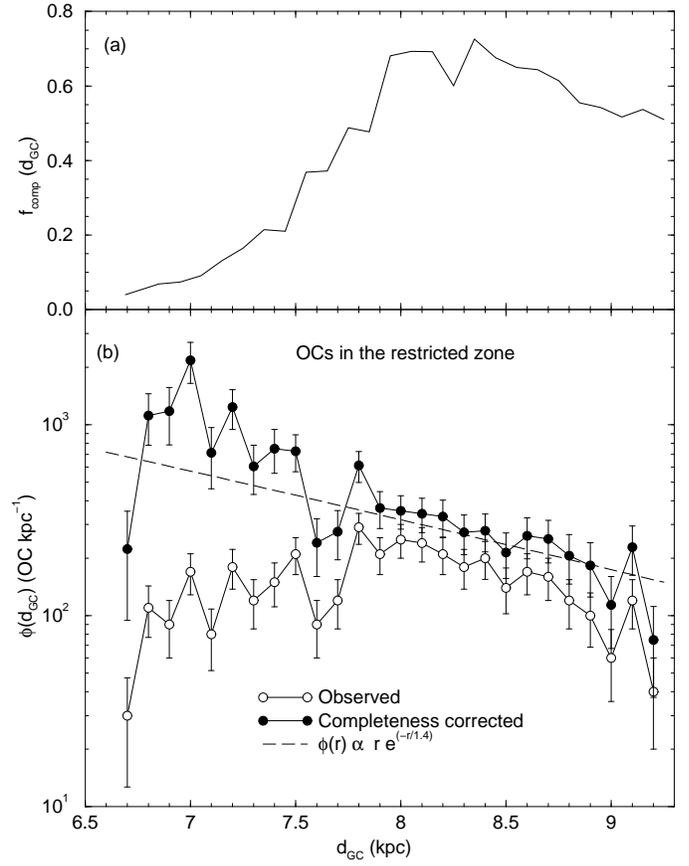}}
\caption[]{Panel (a): completeness fraction as a function of Galactocentric distance in the 
restricted zone. Panel (b): observed (empty circles) and completeness-corrected (filled 
circles) OC distributions. Dashed-line: fit with an exponential-disk with a scale-length 
$\rm R_D=1.4\pm0.2\,kpc$.}
\label{fig11}
\end{figure}

The completeness fraction as a function of Galactocentric distance (for the OCs in the restricted 
zone) is shown in panel (a) of Fig.~\ref{fig11}. As expected, the completeness is highest around the 
Sun ($\rm f_{comp}(8\,kpc)\approx0.68$) and drops off towards external regions and more sharply 
towards the Galactic center. Dividing the observed OC distribution function (Fig.~\ref{fig9}) by 
$\rm f_{comp}(r)$ produces a completeness-corrected radial distribution which follows the
expected exponential-disk profile (panel (b) of Fig.~\ref{fig11}) along the entire Galactocentric 
distance range of the restricted zone, with a scale-length $\rm R_D=1.4\pm0.2\,kpc$. This estimate 
of \rd\ represents about 40\% of the stellar overall disk scale-length (de Vaucouleurs \& Pence 
\cite{deVP78}). Our determination probably reflects the Solar-vicinity value of \rd\ and/or that 
the OCs distribute in the disk with a different scale length than the stars. Remark that the V-shaped 
dip at $\rm\dgc\approx7.5-7.8\,kpc$ (attributed to the edge before the Sgr-Car Arm - Sect.~\ref{SpaDis}) 
is still present in the completeness-corrected distribution function. In this sense, the above
completeness-simulation procedure is sensitive as well to conspicuous physical structures of the 
Galaxy.

Since our analysis is based on magnitude-limited 2MASS photometry ($\rm J=15$), excess 
reddening of background sources towards the inner Galaxy with respect to those in the anti-center 
(Fig.~\ref{fig5}) may have underestimated the contrast parameter of the OCs in that region because 
of the increased differential brightness of the brightest stars. As a consequence, completeness for 
the OCs in the inner Galaxy would be underestimated, which might account for the over-correction of 
$\rm\phi(\dgc)$ with respect to the exponential-disk fit, as suggested by Fig.~\ref{fig11} for 
$\rm\dgc\leq7.5\,kpc$.

Considering the completeness-corrected OC distribution, the fraction of observed OCs in the
restricted zone corresponds to about 47\% of the probable total (essentially of Trumpler types 
I to III - Sect.~\ref{complet}).

\section{Extension to the whole OC sample}
\label{Exten}

In the previous section we corrected for completeness the observed radial distribution of OCs
in the restricted zone and, as a result obtained the expected exponential-disk profile. At this 
point it is interesting to check whether the completeness-simulation procedure retrieves an 
exponential-disk profile when applied to the whole OC sample. We test this hypothesis 
considering 2 different simulation approaches.

\subsection{Actual-position simulation}
\label{APS}

In this simulation the effective number of stars (Sect.~\ref{SimCom}) N* is randomly distributed 
(with the number-frequency according to the histogram in panel (b) of Fig.~\ref{fig10}) among the 
654 OCs assuming the WEBDA spatial locations ($\ell$, $b$, $\ds$). The completeness fraction 
of each OC is then the average value of the retrieval rate after running $10^5$ simulations. Similarly 
to the restricted-zone simulation, the individual $1\sigma$-standard deviation of the completeness 
fraction is at most 3\% of the corresponding average value.

\begin{figure} 
\resizebox{\hsize}{!}{\includegraphics{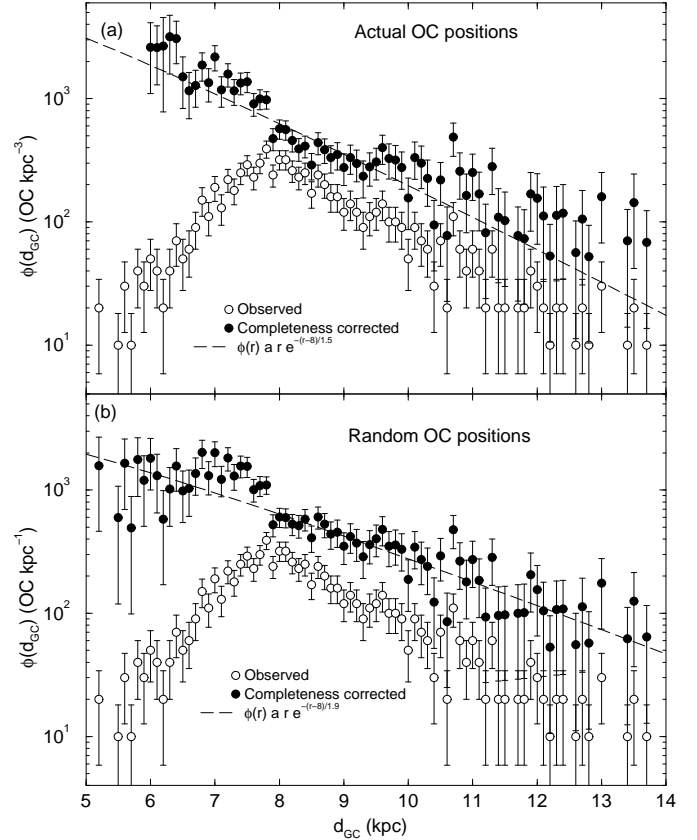}}
\caption[]{Completeness-correction of the observed radial distribution of OCs (empty circles) 
according to the actual-position (top panel) and random-position (bottom panel) simulations. 
The resulting radial distributions (filled circles) follow exponential-disk profiles characterized
by scale-lengths $\rm\rd\approx1.5\,kpc$ and $\rm\rd\approx1.9\,kpc$, respectively.}
\label{fig12}
\end{figure}

The observed radial distribution of OCs in the range $\rm 5\leq\dgc(kpc)\leq14$ (Fig.~\ref{fig12})
presents a maximum at the Solar circle and drops off both towards larger Galactocentric distances 
and more sharply to the Galactic center. The radial completeness fraction resulting from the 
actual-position simulation (Fig.~\ref{fig13}) presents similar features. Correction of the
observed distribution for completeness produces a radial distribution which follows
an exponential-disk profile (panel (a)) characterized by a scale-length of 
$\rm\rd=1.5\pm0.1\,kpc$ (correlation coefficient $\rm CC=0.88$).

The average integrated completeness throughout the disk is $\rm 33.2\pm1.5\%$.

\subsection{Random-position simulation}
\label{RPS}

In this case we build 1000 artificial test positions ($\ell$, $b$, $\ds$) over which we randomly 
distribute the effective number of stars as in the actual-position simulation. The $\ell$ coordinate 
is randomly selected around the Sun while $b$ is drawn according to the observed number-frequency 
histogram shown in panel (e) of Fig.~\ref{fig3}; to match most of the observed distances from the Sun 
(panel (a) of Fig.~\ref{fig4}), \ds\ is randomly selected from the range 0 -- 7\,kpc. Because of the 
random nature of the test OC positions, the resulting radial completeness fraction (Fig.~\ref{fig13}) 
turns out similar to, but considerably smoother than that derived in the actual-position simulation.

The completeness-corrected radial distribution of OCs (panel (b) of Fig.~\ref{fig12}) follows an 
exponential-disk profile characterized by a scale-length $\rm\rd=1.9\pm0.1\,kpc$ ($\rm cc=0.88$), 
about 30\% larger than that obtained with the actual-position simulation, within the uncertainties.

\subsection{Comparison of simulations}

The simulations discussed above are based on conceptually different approaches with respect to
the spatial distribution of OCs. Despite this, they produced similar completeness-corrected 
radial distributions of OCs, which enhances the significant r\^ole played by completeness in 
depth-limited photometric surveys throughout the Galaxy. However, the completeness-corrected
profile obtained with the random-position simulation produces a visually better overall fit 
with the expected exponential-disk profile (Fig.~\ref{fig12}) as compared to the actual-position 
simulation, especially near the inner and outer \dgc\ borders. The completeness-corrected 
distributions derived from both simulation approaches 
present similar disk scale lengths, $\rm\rd\approx1.5-1.9\,kpc$, about 43\% - 54\% of the canonical 
value ($\rm\rd\approx3.5\,kpc$) derived with stars by de Vaucouleurs \& Pence (\cite{deVP78}). 
This fact may reflect intrinsic differences in the spatial distribution of OCs with respect to 
stars.

\begin{figure} 
\resizebox{\hsize}{!}{\includegraphics{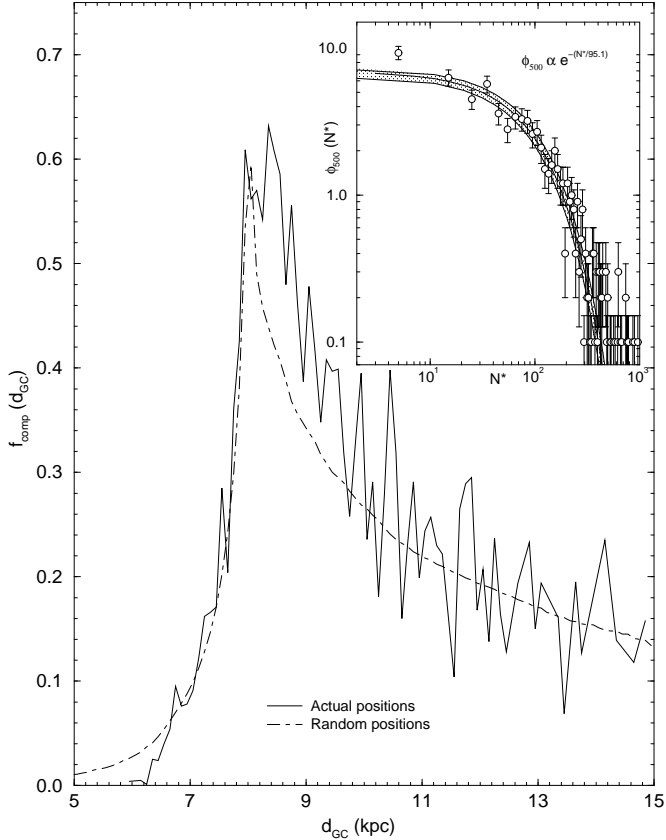}}
\caption[]{The actual-position radial completeness fraction (solid line) drops off more
sharply towards the Galactic center for $\rm\dgc\leq6.6\,kpc$ than that of the random-position
simulation (dot-dashed line). Inset: the distribution of the number of OCs with a given effective 
(background-subtracted) number of stars for the whole OC sample is well fitted by an 
exponential-decay profile. Shaded area: $\rm1\sigma$ standard deviation from the fit.}
\label{fig13}
\end{figure}

In Fig.~\ref{fig13} we compare the radial completeness fractions resulting from both
simulations. They basically agree, except for $\rm\dgc\leq6.6\,kpc$ where the actual-position
completeness fraction drops off more sharply towards the Galactic center than that from
the random-position simulation. Considering that in the random-position simulation
the OCs are uniformly distributed throughout the Galaxy, as opposed to observed OC
coordinates in the actual-position, perhaps this difference may suggest that the 
tidal-disruption effects due to the Galactic center/bulge begin to be observationally 
significant at $\rm\dgc\approx6.6\,kpc$. 

Finally, in the inset of Fig.~\ref{fig13} we show the distribution of the number of OCs
with a given effective (background-subtracted) number of stars for the whole OC sample. 
Similarly to the OCs in the restricted zone this distribution follows an exponential-decay 
profile, but with a characteristic number of stars of $\Delta N*\approx95$. The decay-scale 
for the whole sample is about 4 times larger than that derived for the OCs outside the Solar 
circle in the restricted zone (Sect.~\ref{SimCom}). This discrepancy can be accounted for by 
an observational bias, since detection probability in directions intercepting the Galactic 
center/bulge is higher for the more populous OCs. Interestingly, the analytical expression of 
the overall $\rm\phi_{500}$ resembles that of the galaxy luminosity function (Schechter 
\cite{Sch76}).

\subsection{Completeness in vertical distributions}

In Sect.~\ref{DepenZH} we showed that the observed scale height depends on Galactocentric
distance, in the sense that the OC distribution outside the Solar circle is about $\rm2\times$ 
broader than that inside it. At this point it may be interesting to determine how completeness 
affects the vertical distribution and derive its intrinsic shape. 

\begin{figure} 
\resizebox{\hsize}{!}{\includegraphics{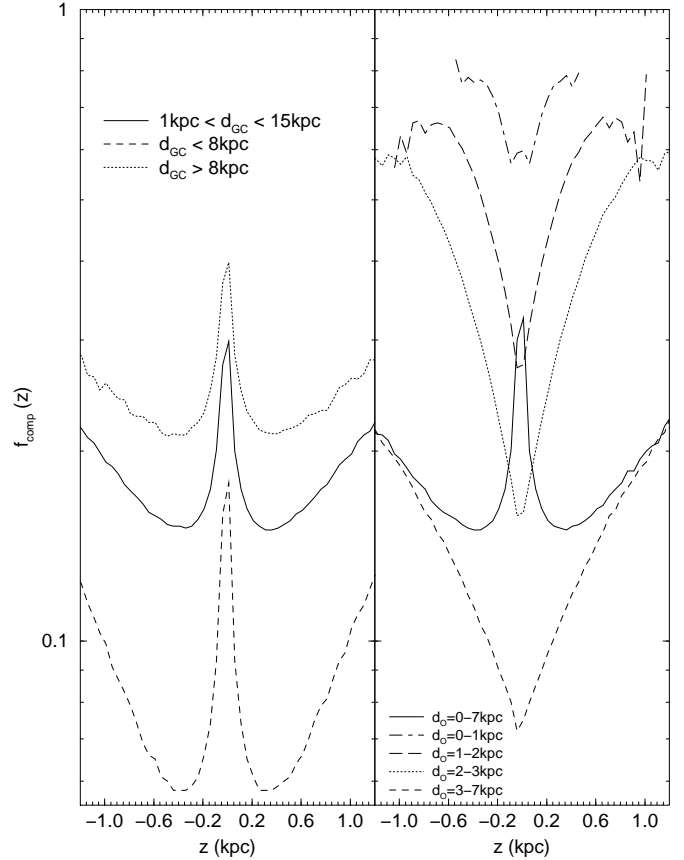}}
\caption[]{Completeness fraction as a function of z. Left panel: completeness fraction for OCs
internal (dashed line) and external (dotted line) to the Solar circle, and the average disk
completeness (solid line). Right panel: the shape of the average disk completeness results from the
combination of OCs at different ranges of distance from the Sun.}
\label{fig14}
\end{figure}

In Fig.~\ref{fig14} (left panel) we show how the completeness fraction as a function of z depends 
on Galactocentric distance, centered at the Solar position over the disk ($\rm\zo=14.81\,pc$, 
Sect.~\ref{dsh}). The curves have been produced with the random-position simulation (Sect.~\ref{RPS}).
As expected, the OCs internal to the Solar circle suffer more severely from 
completeness than the external ones. Interestingly, both curves present similar shapes, with a peak 
at $\rm z=\zo$, sharp declines for increasing $\rm|z|$ with a minimum at $\rm|z|\approx340\,pc$,
and an increase for larger values of $\rm|z|$. As shown in the right panel OCs at different ranges
of distance from the Sun present similar dependence with z, with a minimum at $\rm z=\zo$ and an
increase for larger $\rm|z|$. However, the absolute values of completeness drop for more distant
OCs. The final shape of the completeness fraction results from the combination of OCs at different 
distances from the Sun. In particular, the peak at $\rm z=\zo$ can be accounted for by the high 
completeness of the OCs closer than $\rm\approx1\,kpc$ from the Sun. 

\begin{figure} 
\resizebox{\hsize}{!}{\includegraphics{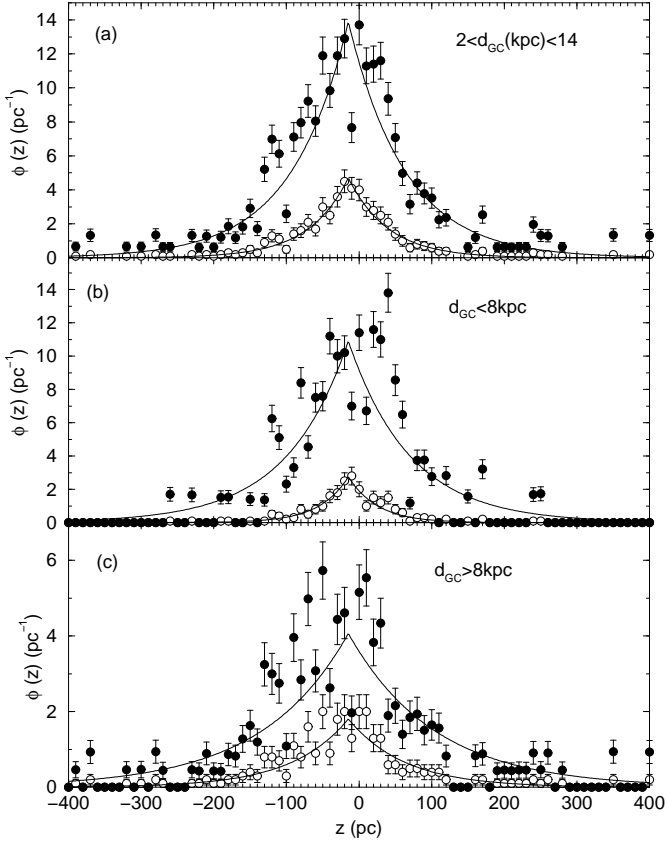}}
\caption[]{Completeness-corrected z-distributions. Panel (a): overall radial distribution (scale 
height $\rm\zh\approx78\,pc$). Panel (b): OCs internal to the Solar circle ($\rm\zh\approx85\,pc$). 
Panel (c): OCs external to the Solar circle ($\rm\zh\approx119\,pc$). Empty circles: 
observed OCs. Filled circles: completeness-corrected z-distribution.}
\label{fig15}
\end{figure}

The observed z-distributions of OCs internal and external to the Solar circle, as well as
the overall one have been corrected for completeness with the corresponding curves derived 
above. The resulting completeness-corrected distributions along with the respective
exponential-decay fits are shown in Fig.~\ref{fig15}. The completeness-corrected scale heights 
are $\rm\zh=78\pm3\,pc$ (observed: $\rm\zh\approx57\,pc$) with correlation coefficient $\rm 
CC=0.88$, $\rm\zh=85\pm13\,pc$ (observed: $\rm\zh\approx39\,pc$) with $\rm CC=0.77$, and 
$\rm\zh=119\pm12\,pc$ (observed: $\rm\zh\approx78\,pc$) with $\rm CC=0.75$, respectively for 
the overall, internal and external to the Solar circle 
distributions. Consequently, completeness correction of the z-distributions preserved the 
dependence of scale height on Galactocentric distance, increased the absolute values of \zh\ 
with respect to the observed ones and decreased the observed ratio of \zh\ outside/inside the 
Solar circle from 2 to 1.4. 

Completeness-correction of the observed z-distributions of the OCs younger than 200\,Myr
and those with age in the range 200\,Myr -- 1\,Gyr (panels (b) and (d) of Fig.~\ref{fig7},
respectively) increased the scale height by a factor of $\sim1.42$. Consequently, the young
OCs distribute vertically following an exponential-decay profile with scale height 
$\rm\zh\approx68\,pc$. For the 200\,Myr -- 1\,Gyr OCs the completeness-corrected scale height
is $\rm\zh\approx230\,pc$. Considering the observational uncertainties, these values are 
comparable to the \ion{H}{i} scale height in the Solar neighbourhood, which we take as 
$\rm\sim\frac{1}{2}FWHM(\ion{H}{i})\approx115\,pc$ (Dickey \& Lockman \cite{DL90}). This must 
reflect the association of star formation and the parent gas. The same conclusion applies to 
CO, for which surveys of the outer Galaxy (Heyer et al. \cite{Heyer98}), where the Perseus arm 
lies, find an overall mean CO scale height of 113\,pc.

\subsection{Age distribution function}
\label{tau_CC}

\begin{figure} 
\resizebox{\hsize}{!}{\includegraphics{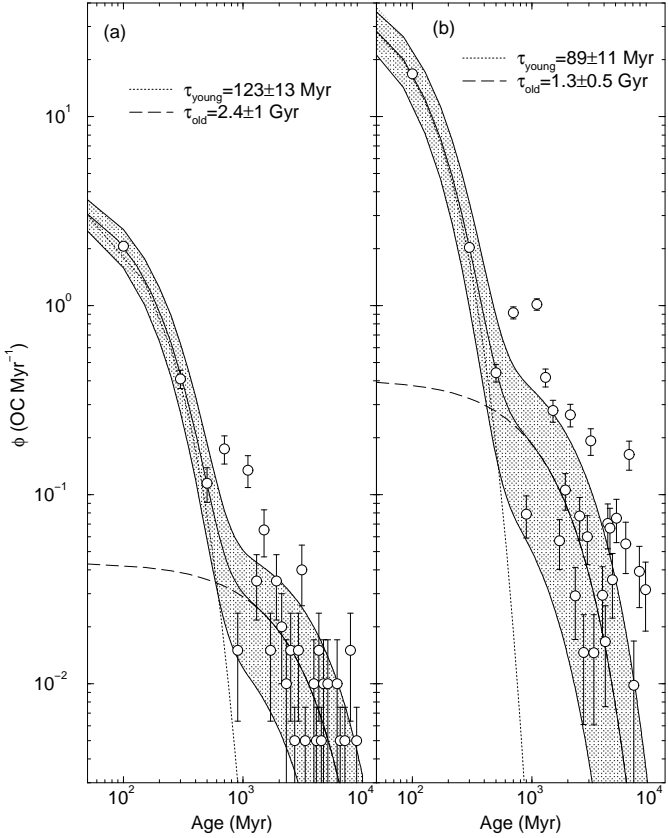}}
\caption[]{Panel (a): the observed age-distribution can be fitted with 
two exponential-decay laws with time-scales of $\rm\tau_{young}=123\pm13\,Myr$ (dotted line) and 
$\rm\tau_{old}=2.4\pm1\,Gyr$ (dashed line) characterizing the young and old OCs, respectively.
Panel (b): same as (a) for the completeness-corrected age distribution; time-scales are
$\rm\tau_{young}=89\pm11\,Myr$ (dotted line) and $\rm\tau_{old}=1.3\pm0.5\,Gyr$ (dashed line).
Shaded region: $\rm1\sigma$-standard deviation of the combined young$+$old fit.}
\label{fig16}
\end{figure}

The histogram with the number of observed OCs in bins of age (Fig.~\ref{fig2}) was converted into 
the observed age-distribution function (panel (a) of Fig.~\ref{fig16}). This function can be fitted 
by a combination of exponential-decay profiles with age scales of $\rm\tau_{young}=123\pm13\,Myr$
(characteristic of the young OC population) and $\rm\tau_{old}=2.4\pm1\,Gyr$ (old OCs). 
The correlation coefficient of the fit is $\rm CC=0.94$. Clearly the observed old OC population
represents an excess over the extrapolation to old ages of the young OCs distribution function.
An additional excess in the observed age-distribution function shows up in the age range 
$\rm\sim600\,Myr\ to\ \sim1.5\,Gyr$, which can probably be identified with the moderate-age OC 
population. The present young and old age time-scales are comparable to those derived by Janes \& Phelps 
(\cite{JP94}), 200\,Myr and 4\,Gyr for the young and old OCs, respectively.

The observed age-distribution function (panel (a) of Fig.~\ref{fig16}) basically contains the 
survivor clusters after emerging from the parent molecular clouds. Lada \& Lada (\cite{lada03}) 
estimate that, for embedded clusters within 2\,kpc from the Sun, only a fraction of 4-7\% emerge 
as OCs. Bica, Dutra \& Barbuy (\cite{BiDuBa03}) compiled 276 embedded clusters (mostly infrared 
clusters) and stellar groups from the literature located mostly within 3\,kpc from the Sun. From 
the age-distribution function in Fig.~\ref{fig16} we estimate that the number of observed OCs 
with ages in the ranges 5--10\,Myr, 10--15\,Myr, and 15--20\,Myr are $21\pm3$, $21\pm3$ and 
$20\pm3$, respectively. Considering an age range of 5\,Myr for the embedded clusters (typical 
lifetime of \ion{H}{ii} regions) and comparing with the number of OCs in the subsequent 5\,Myr-bins 
(see above) we can derive a 
survival rate of $\sim8\%$, similar to Lada \& Lada (\cite{lada03})'s upper limit. More recently 
Dutra et al. (\cite{DBSB03}) and Bica et al. (\cite{BDSB03}) surveyed for new embedded clusters,
groups and candidates in the direction of radio and optical \ion{H}{ii} regions and molecular clouds. 
These objects are typically within $\sim5$\,kpc from the Sun according to kinematical distances 
of the related \ion{H}{ii} regions. Adding these newly discovered embedded clusters to those of Bica, 
Dutra \& Barbuy (\cite{BiDuBa03}), totaling 622 objects, the survival rate drops to $\approx3.4\%$. 
The present estimates match those of Lada \& Lada (\cite{lada03}).

One output of the actual-position simulation (Sect.~\ref{APS}) is the average individual 
completeness of each OC in the present sample. The reciprocal of the completeness is essentially 
the detection probability of that particular OC - and consequently of its intrinsic parameters, 
such as the age. In this sense we can apply the same procedure to correct the observed age 
distribution for completeness. The resulting completeness-corrected age distribution (panel (b) 
of Fig.~\ref{fig16}) retains the basic features of the observed one, including the 
intermediate-age OC excess. However, because the distribution in Galactic latitude of the young 
OCs is narrower than that of the old OCs (Fig.~\ref{fig3}), the young OCs are subject on average 
to higher incompleteness. Consequently, completeness correction should enhance the number of young 
OCs relative to the old ones. Indeed, the observed ratio of the number of OCs younger than 1\,Gyr 
with respect to the older ones is $\approx7$, while in the completeness-corrected distribution 
this value increases to $\approx14$.

Similarly to the observed profile, the best-fit of the completeness-corrected age distribution 
results from a combination of exponential-decay curves with time scales $\rm\tau_{young}=89\pm11\,Myr$ 
and $\rm\tau_{old}=1.3\pm0.5\,Gyr$ ($\rm CC=0.94$), respectively for the young and old OCs. The 
completeness-corrected time scales result comparable to the observed ones, within the uncertainties.
We note however that the resulting time scales of both OC populations correspond to about half of 
the corresponding values estimated by Janes \& Phelps (\cite{JP94}). This might suggest faster
destruction-rates than those implied by Janes \& Phelps (\cite{JP94}), and particularly with
respect to the $\rm\sim600\,Myr$ destruction-time scale suggested by Bergond, Leon \& Guibert 
(\cite{Bergond2001}).

From the above we conclude that evolutionary scenarios based on constant rates of cluster formation 
and disruption do not apply to the double exponential-decay age distribution. Probably a 
time-varying star-formation rate, such as a massive initial burst at the Gyr scale, might fit the
data. In this context, the suggestion of a galaxy merger in the early phases of the Milky Way (Chen, 
Stoughton, Smith et al. \cite{ChenSS01}) and the subsequent disturbances to the disk and burst of
star formation, could account for the observed excess in the number of old OCs surviving to the
present time. One clue to disentangle the cluster formation/disruption-rate problem may be provided 
by the systematic study of OC remnants (e.g. Pavani et al. \cite{Pavani03}).

\subsection{Galactocentric distance {\it vs.} age}
\label{dgcXage}

\begin{figure} 
\resizebox{\hsize}{!}{\includegraphics{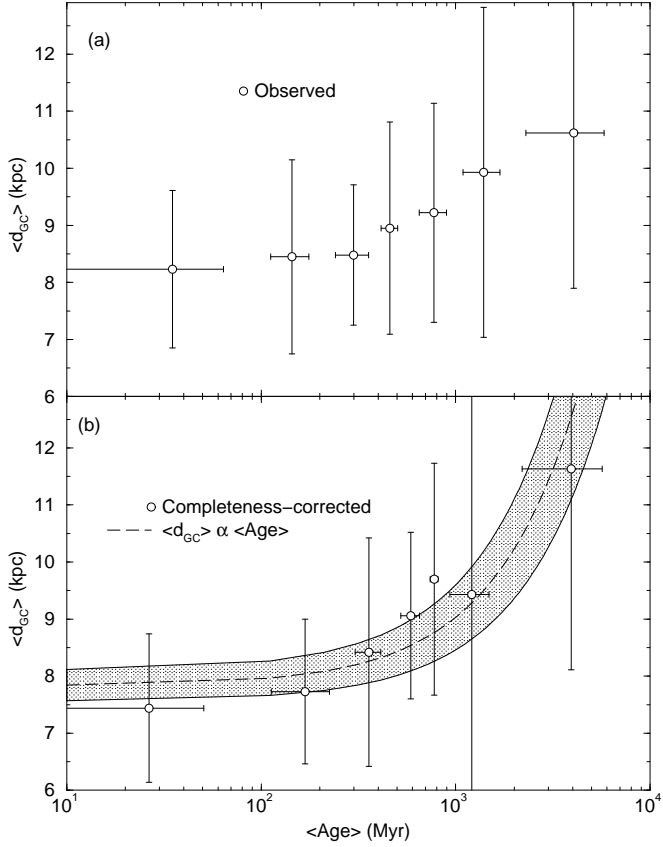}}
\caption[]{Relation of Galactocentric distance and age. Panel (a): observed OCs. Panel (b):
completeness-corrected. The bars show the $\rm1\sigma$-standard deviations of the average 
values of age and Galactocentric distance within the respective bins. Dashed line: linear-least 
squares fit to the completeness-corrected points. Shaded region: $\rm1\sigma$-standard
deviation of the fit.}
\label{fig17}
\end{figure}

In Fig.~\ref{fig17} we examine the relation of Galactocentric distance with age. We divided the
OC sample in bins of age and calculated the average age and \dgc\ in each bin, as well as the 
respective standard deviations. Lyng\aa\ (\cite{Lynga82}) and Tadross et al. (\cite{Tad2002}) 
concluded that older clusters are found preferentially in the outer parts of the Galaxy, while 
younger clusters are more evenly distributed. The observed OC distribution is in panel (a) and 
the completeness-corrected one is in panel (b). The large $\rm1\sigma$-standard deviation bars 
in the observed distribution, particularly in \dgc, preclude any definitive conclusion on 
this issue. 

However there appears to exist a trend in which older clusters are found preferentially at larger 
Galactocentric distances. This trend is enhanced after applying the completeness correction to
the observed distribution (panel (b)). A linear-least squares fit to the completeness-corrected
distribution results in the relation $\rm\langle\dgc\rangle=(7.83\pm0.27)+(0.0012\pm0.0003)
\langle Age\rangle$ ($\rm CC=0.87$), for \dgc\ in kpc and Age in Myr. 

\section{Estimate of the total number of open clusters}
\label{tocg}

We describe the spatial number-density of open clusters in the disk as a combination of 
exponential-decay laws for the radial and vertical components,
$\rm\rho(r,z)=\frac{dN_{oc}}{rdrdz\,d\theta}=\rho_o\,e^{-(r/R_D)}e^{-|z/z_h|}$, where $\rm R_D$ 
is the disk scale-length, \zh\ is the scale-height and $\rm\rho_o$ is the number-density of OCs 
at the Galactic center. 
Using cylindrical coordinates and defining $\rm D_\odot$ as the Galactocentric distance of the 
Sun, \zo\ the displacement of the Sun from the Galactic plane, and \po\ the 
local (solar vicinity) number-density of OCs, we can write 
$\rm\rho(r,z)=\rho_\odot e^{-\left(\frac{r-D_\odot}{R_D}\right)}
e^{-\left(\frac{|z|-z_\odot}{z_h}\right)}$. Integration over the z-axis and through a region
with opening angle $\Delta\theta$ produces the radial distribution function of OCs
\begin{equation}
\label{eq1}
\rm\phi(r)=\frac{dN_{oc}}{dr}=4\pi\left(\frac{\Delta\theta}{2\pi}\right)\rho_\odot\,r\,z_h 
e^{\frac{z_\odot}{z_h}}e^{-\left(\frac{r-D_\odot}{R_D}\right)}. 
\end{equation}
The total number of open clusters in the disk is then ($\Delta\theta=2\pi$)
\begin{equation}
\label{eq2}
\rm N_{oc}=4\pi\rho_\odot\zh\,e^{\frac{z_\odot}{z_h}}e^{\left(\frac{D_\odot}{R_D}\right)}R_D^2 
\left[1-\left(1+\frac{R_G}{R_D}\right)e^{-\left(\frac{R_G}{R_D}\right)}\right],
\end{equation}
where $\rm R_G$ is the disk radius ($\rm R_G\approx15\,kpc$, Binney \& Tremaine 
\cite{BinTre1987})\footnote{Because of the exponential factor in eq.~3 $e^{-\left(\frac{R_G}{R_D}\right)}$
with $\rm R_D=1.5-1.9\,kpc$ (Sects.~\ref{APS} and \ref{RPS}), our estimates of $\rm N_{oc}$ are 
essentially insensitive to values of $\rm R_G\geq10\,kpc$.}. 
Accordingly, estimates of both the solar-vicinity OC density and the local disk scale length can 
be obtained by fitting eq.~\ref{eq1} to completeness-corrected distribution functions.

Based on eq.~\ref{eq1} and the completeness-corrected distribution functions of the actual-position 
and random-position simulations for the whole OC sample (with $\rm\frac{\Delta\theta}{2\pi}\approx0.11$, 
Fig.~\ref{fig1}) we derive $\rm\po=759\pm34\,kpc^{-3}$ and $\rm\po=774\pm33\,kpc^{-3}$, respectively
(Fig.~\ref{fig12}).

Extrapolation of the above values of \po\ and \rd\ to the whole disk using eq.~\ref{eq2}
produces a total number of OCs in the Galaxy of $\sim3.7\times10^5$ and $\sim1.8\times10^5$,
respectively for the actual-position and random-position completeness-corrected distribution 
functions. These estimates were obtained by direct extrapolation of the completeness-corrected 
distributions down to the Galactic center. Consequently, they do not take into account depletion 
in the number of OCs by tidal disruption and enhanced frequency of collisions with molecular 
clouds in the inner parts of the Galaxy. In this sense, the above numbers of Galactic OCs must 
be taken as upper-limits.

For the completeness-corrected distribution function in the restricted zone (with a scale-height 
of 57\,pc (Table~\ref{tab1}) and $\rm\frac{\Delta\theta}{2\pi}=0.05$) we derive 
$\rm\rd=1.4\pm0.2\,kpc$ and $\rm\po=795\pm70\,kpc^{-3}$ (Fig.~\ref{fig11}). This number can be 
compared with that obtained assuming a uniform, cylindrically-symmetric distribution of OCs around 
the Sun. With a total number of OCs up to a distance from the Sun of 1.3\,kpc of 341, and using a 
vertical height of $\approx2\times$ the scale-height we derive $\rm\po=281\pm16\,kpc^{-3}$, which 
corresponds to the observed OCs. Correcting for the average completeness in the restricted zone 
($\sim47\%$) produces a probable total of $\sim730$ OCs, and a number density of $\rm\po=600\pm33\,kpc^{-3}$, 
which within the uncertainties roughly agrees with the value derived from the distribution function.

\section{Concluding remarks}
\label{Conclu}

In the present paper we use a sample of 654 open clusters (typically Trumpler types I to III) 
with published parameters to derive statistical properties related to age, distance from the Sun, 
reddening, Galactocentric distance and height with respect to the Galactic plane. 

The population of OCs younger than 200\,Myr distributes vertically in the disk following an 
exponential-decay profile with a scale height of $\zh=47.9\pm2.8$\,pc. Clusters with ages in 
the range 200\,Myr to 1\,Gyr distribute vertically with $\zh=150\pm27$\,pc, while older cluster 
distribute nearly uniformly in height from the plane so that no scale height can be derived from 
exponential fits. The average scale height, considering clusters of all ages, is $\zh=57.2\pm2.8$\,pc. 
The above OC scale heights are considerably smaller than those attributed to the thin disk 
($\rm\zh\approx0.6\,kpc$) as derived by means of stars (e.g. de Boer et al. \cite{BAA97};
Altmann et al. \cite{AEB04}; Kaempf, de Boer \& Altmann \cite{KBA05}). We confirm previous
findings that \zh\ increases with Galactocentric distance, being about twice as large in regions 
outside the Solar circle than inside it.

The asymmetry in the vertical distribution of OCs allowed us to derive the displacement of the 
Sun above the Galactic plane as $\rm\zo=14.8\pm2.4$\,pc, which agrees with previous determinations 
using stars. 

Sample completeness affects critically the detection particularly of low-contrast OCs internal to 
the Solar circle and those distant from the Sun. We simulate the effects of completeness in
all directions and for any distance from the Sun assuming that the intrinsic distribution of the 
number of OCs with a given number of stars (above the background) measured in a restricted zone 
outside the Solar circle is isotropic. As a consequence we derived the radial dependence of
completeness. The observed radial distribution of OCs in terms of Galactocentric distance does not 
follow the expected exponential profile, instead it falls off both for regions external to the 
Solar circle and more sharply towards the Galactic center. Correction for completeness produced 
radial distributions which agree with exponential disks throughout the Galactocentric distance 
range 5--14\,kpc, with scale lengths $\rm\rd=1.5 - 1.9\,kpc$. These values of \rd\ are smaller 
than those implied by stars, which may reflect intrinsic differences in the spatial distribution 
of OCs and stars. 

With respect to the vertical disk structure described by the OCs, we showed that older clusters 
distribute in exponential disks with larger scale heights than young OCs. In addition we confirmed
previous observations that the scale height increases with Galactocentric distance. The latter result 
was further enhanced by the completeness-corrected z-distributions. The completeness-corrected OC 
scale height is comparable to those implied both by the \ion{H}{i} and CO distributions.

We derived a number-density of solar-neighbourhood (with distances from the Sun $\rm\ds\leq1.3\,kpc$) 
OCs of $\rm\po=795\pm70\,kpc^{-3}$, which implies a total number of (basically Trumpler types I to III) 
OCs of $\sim730$ of which $\sim47\%$ would already have been observed. Inferences on the total 
number of OCs in the disk can be made based on the local number density and radial/azimuthal 
completeness-corrected OC distributions. This estimate is important for the realization of what to 
expect in future surveys of the Galaxy. Direct extrapolation of the completeness-corrected 
distributions down to the Galactic center suggests that the total number of OCs in the Galaxy is in 
the range $\rm(1.8 - 3.7)\times10^5$. These numbers must be taken as upper-limits because the
extrapolation does not take into account depletion in the number of OCs by tidal disruption and 
enhanced frequency of collisions with molecular clouds in the inner parts of the Galaxy.

The observed and completeness-corrected OC age-distribution functions can be described 
by a combination of two exponential-decay profiles characterizing the young and old OC 
populations with age scales of $\rm\sim100\,Myr$ and $\rm\sim1.9\,Gyr$, respectively. As
a consequence, scenarios based on constant star-formation and OC-disruption rates do not
apply to the data. Comparing the number of embedded clusters from IR studies in the literature 
and the present observed age-distribution function of very young OCs we derive survival rates 
in the range 3.4--8\%, in agreement with the estimates of Lada \& Lada (\cite{lada03}).

\begin{acknowledgements}
We thank an anonymous referee for interesting remarks.
This publication makes use of the WEBDA open cluster database, as well as data products from 
the Two Micron All Sky Survey, which is a joint project of the University of Massachusetts and 
the Infrared Processing and Analysis Center/California Institute of Technology, funded by the 
National Aeronautics and Space Administration and the National Science Foundation. 
We acknowledge support from the Brazilian Institution CNPq.
\end{acknowledgements}

%


\begin{thebibliography}{}


\bibitem[2004]{AEB04} 
   Altmann, M., Edelmann, H. \& de Boer, K.S. 2004, A\&A, 414, 181

\bibitem[1981]{BS81}
   Bahcall, J.N. \& Soneira, R.M. 1981, ApJS, 47, 357

\bibitem[2001]{Bergond2001}
   Bergond, G., Leon, S. \& Guibert, J. 2001, A\&A, 377, 462

\bibitem[2003]{BiBoDu03}
   Bica, E., Bonatto, C. \&  Dutra, C.M. 2003, A\&A, 405, 991
   
\bibitem[2005a]{BiBo2005a}   
    Bica, E. \& Bonatto, C.J. 2005a, A\&A, 431, 943
   
\bibitem[2005b]{BiBo2005b}   
    Bica, E. \& Bonatto, C.J. 2005b, A\&A, submitted
    
\bibitem[2003]{BiDuBa03}    
    Bica, E., Dutra, C.M. \& Barbuy, B.  2003, A\&A, 397, 177
    
\bibitem[2003]{BDSB03}  
   Bica, E., Dutra, C.M., Soares, J. \&  Barbuy, B. 2003, A\&A, 404, 223   

\bibitem[1987]{BinTre1987}
   Binney, J. \& Tremaine, S. 1987, in {\it Galactic Dynamics}, Princeton,
   NJ: Princeton University Press. (Princeton series in astrophysics)


\bibitem[1997]{BAA97} 
   de Boer, K.S., Aguilar Sanchez, Y. \& Altmann, M. et al. 1997, A\&A, 327, 577

\bibitem[2004]{BoBiDu04} 
   Bonatto, C., Bica, E. \& Dutra, C.M. 2004, A\&A, 422, 555

\bibitem[2005]{BB2005} 
   Bonatto, C.J. \& Bica, E. 2005, A\&A, in press.

\bibitem[2001]{ChenSS01}
   Chen, B., Stoughton, C., Smith, J.A. et al. 2001, ApJ, 553, 184

\bibitem[1995]{Cohen95}
   Cohen, M. 1995, ApJ, 444, 874
   
\bibitem[1978]{deVP78}
   de Vaucouleurs, G. \& Pence, W.D. 1978, AJ, 83, 1163   

\bibitem[2002]{Dias2002}
   Dias, W.S., Alessi, B.S., Moitinho, A. \& L\'epine, J.R.D. 2002, A\&A, 389, 871

\bibitem[1990]{DL90}
   Dickey, J.M. \& Lockman, F.J. 1990, ARA\&A, 28, 215

\bibitem[2003]{DBSB03} 
    Dutra, C.M., Bica, E., Soares, J. \& Barbuy, B. 2003, A\&A, 400, 533
    
\bibitem[1976]{Faber76}
   Faber, S.M., Burstein, D., Tinsley, B. \& King, I.R. 1976, AJ, 81, 45

\bibitem[1995]{Friel95}
   Friel, E.D. 1995, ARA\&A, 33, 381

\bibitem[1998]{delaF98}
   de la Fuente Marcos, R. 1998, A\&A, 333, L27

\bibitem[1970]{GG70}
   Georgelin, Y.P. \& Georgelin, Y.M. 1970, A\&A, 6, 349
   
\bibitem[2002]{Girardi2002} 
   Girardi, L., Bertelli, G., Bressan, A., et al. 2002, A\&A, 391, 195

\bibitem[1995]{HGMC95}
   Hammersley, P.L., Garz\'on, F., Mahoney, T. \& Calbet, X. 1995, 
   MNRAS, 273, 206
   
\bibitem[1998]{Heyer98}
   Heyer, M.H., Brunt, C., Snell, R.L., Howe, J., Schloerb, F.P. \& Carpenter, J.M. 
   1998, ApJS, 115, 241

\bibitem[1982]{JA82}  
   Janes, K. \& Adler, D. 1982, ApJS, 49, 425
   
\bibitem[1994]{JP94}  
   Janes, K. \& Phelps, R.L. 1994, AJ, 108, 1773

\bibitem[2005]{KBA05}
   Kaempf, T.A., de Boer, K.S. \& Altmann, M. 2005, A\&A, 432, 879
   

\bibitem[1991]{KDF91}
   Kent, S.M., Dame, T.M. \& Fazio, G. 1991, ApJ, 378, 131

\bibitem[1966]{King1966}
   King, I. 1966, AJ, 71, 64
   
\bibitem[1978]{Knapp78}
   Knapp, G.R., Tremaine, S.D. \& Gunn, J.E. 1978, AJ, 83, 1585

\bibitem[2003]{lada03}
   Lada, C.J. \& Lada, E.A. 2003, ARA\&A, 41, 57  
      
\bibitem[1982]{Lynga82}
   Lyng\aa, G. 1982, A\&A, 109, 213
   
\bibitem[1993]{Maj93} 
   Majewski, S.R. 1993, ARA\&A, 31, 575  
   
\bibitem[1969]{McC69}
   McCuskey, S.W. 1969, AJ, 74, 807
      
\bibitem[1996]{Merm1996}   
   Mermilliod, J.C. 1996, in {\it The Origins, Evolution, and Destinies of 
   Binary Stars in Clusters}, ASP Conference Series, eds. E.F. Milone \& J.-C. 
   Mermilliod, 90, 475
   
\bibitem[2002]{Nilakshi2002} 
   Nilakshi, S.R., Pandey, A.K. \& Mohan, V. 2002, A\&A, 383, 153

\bibitem[2005]{OBB05}
   Ortolani, S., Bica, E. \& Barbuy, B., 2005, A\&A, in press

\bibitem[2003]{Pavani03}
   Pavani, D.B., Bica, E., Ahumada, A.V. \& Clari\'a, J.J. 2003, A\&A, 399, 113
         
\bibitem[1998]{Platais98}
   Platais, I., Kozhurina-Platais, V. \& van Leeuwen, F. 1998, AJ, 116, 2423

\bibitem[1993]{Reid93}
   Reid, M.J. 1993, ARA\&A, 31, 345 

\bibitem[1966]{Rup66}
   Ruprecht, J. 1966, BAICz, 17, 33
   
\bibitem[1955]{Salp55}
   Salpeter, E. 1955, ApJ, 121, 161
   
\bibitem[1976]{Sch76}
   Schechter, P. 1976, ApJ, 203, 297
   
\bibitem[1997]{2mass1997} 
   Skrutskie, M., Schneider, S.E., Stiening, R., et al. 1997, in {\it The Impact 
   of Large Scale Near-IR Sky Surveys}, ed. Garzon et al., Kluwer (Netherlands), 210, 187
   
\bibitem[2002]{Tad2002}
   Tadross, A.L., Werner, P., Osman, A. \& Marie, M. 2002, NewAst, 7, 553

   
   
\bibitem[1980]{vdBM80}
   van den Bergh, S. \& McClure, R.D. 1980, A\&A, 88, 360
   
\end{thebibliography}
\end{document}